\documentclass[final,3p,times,authoryear]{elsarticle}

\usepackage{amsmath, amssymb, amsfonts, amsthm}  
\usepackage{graphicx}          
\usepackage{siunitx}           
\usepackage{bm}
\usepackage{upgreek}
\usepackage{breakurl}
\usepackage{pdfpages}
\usepackage[hidelinks]{hyperref}
\usepackage{rotating}
\usepackage{appendix}

\begin{document}

\begin{frontmatter}
    \title{Frequency-dependent stress response under thermal cycle: A thermal-crystal plasticity and dynamic mode decomposition study
    \footnote{This is the accepted manuscript of an article published in \textit{International Journal of Plasticity}. The final published version is available at \url{https://doi.org/10.1016/j.ijplas.2026.104722}.}
    }

    \author[1]{Haruki Ohashi
    }
    \ead{haruki.ohashi.p2@dc.tohoku.ac.jp}
    
    \author[1]{Yoshiteru Aoyagi\corref{cor1}
    }
    \ead{aoyagi@tohoku.ac.jp}

    \cortext[cor1]{Corresponding author}

    \affiliation[1]{
        organization={Department of Finemechanics, Tohoku University},
        addressline={6-6-01 Aoba, Aramaki, Aoba-ku},
        city={Sendai},
        country={Japan}
    }

    \begin{abstract}
        Thermal cycle environments involving repeated temperature changes are common conditions observed in engine components, electronic parts, and additive manufacturing processes. Under such conditions, materials undergo repeated thermal expansion and contraction, forming complex thermal stress fields. Thermal-crystal plasticity simulations that account for stress fields and thermal conduction at the polycrystalline microstructure scale are an effective method for numerically reproducing thermal cycle environments and individually evaluating the influence of factors that are difficult to control experimentally. However, the influence of thermal cycle frequency on the temporal behavior of the stress field and plastic response has not yet been fully understood, partly because a systematic analysis method capable of simultaneously capturing spatial heterogeneity and temporal evolution remains limited.
        In this study, we predicted the polycrystalline-scale thermal stress field generated under different thermal cycle frequencies using thermal-crystal plasticity finite element simulations and investigated the effect of frequency on the spatiotemporal structure of the stress response. 
        The results revealed that under low-frequency conditions, the stress response exhibits quasi-steady-state behavior synchronized with the thermal cycle. 
        Conversely, under high-frequency conditions, the response becomes increasingly inharmonic and non-stationary. 
        While this transition is qualitatively consistent with expectations based on the Fourier number, the present framework further illustrate that the resulting thermal-mechanical response can be represented as a superposition of multiple effective temporal components, reflecting the increased complexity of the system behavior.
        By employing dynamic mode decomposition (DMD) as a diagnostic and post-processing technique, we demonstrate that the spatiotemporal structure of the stress field under thermal cycle conditions can be systematically extracted and compactly represented. This approach enables a quantitative characterization of frequency-dependent changes in the thermal stress response beyond conventional averaging or snapshot-based analyses. The results highlight the utility of DMD as a framework for organizing complex simulation data and for interpreting the temporal structure of plastic response under cyclic thermal loading.
    \end{abstract}

\end{frontmatter}

\section{Introduction}
Thermal loading environments are ubiquitous in modern engineering applications, including high-temperature structural components \citep{Skamniotis_etal_2023,Alipooramirabad_etal_2024,Culafic_etal_2025}, additive manufacturing processes \citep{Li_etal_2018,Bronkhorst_etal_2019,Kuang_etal_2026}, and electronic devices \citep{Yilun_etal_2022,Wang_etal_2024,Qian_etal_2024}.
Under these conditions, thermal stress and residual stress arise within the material due to temperature changes, and their distribution and residual patterns significantly affect mechanical properties and service life \citep{Peterson_etal_2024,Ren_etal_2025,Roy_etal_2025}.
Underestimating residual stress fields can lead to failure and accidents, while accurately understanding and controlling them enables designs that maximize a material's inherent performance \citep{Tabatabaeian_etal_2022,Bandyopadhyay_etal_2024}.
Furthermore, the thermal fatigue phenomenon caused by repeated thermal loading is an unavoidable challenge for component life prediction and reliability design.

To quantitatively understand the stresses and deformations associated with thermal fatigue, it is essential to perform coupled analysis integrating heat conduction with mechanical responses at the grain scale.
Thermal fatigue involves complex grain-scale mechanics and thermodynamics, making experimental control of microstructure and temperature field challenging.
As a framework for such analysis, coupled thermal-crystal plasticity analysis, which combines heat conduction and crystal plasticity analyses \citep{Han_etal_2020,Sedighiani_etal_2021}, has been widely used in recent years.
Coupled thermal-crystal plasticity analysis enables high-resolution prediction of stress and strain field evolution under thermal loadings while accounting for anisotropic elastic and plastic responses within polycrystalline structures and deformation behavior that depends on crystal orientation \citep{Roters_etal_2019}.
Existing research has thoroughly discussed the interaction between thermal effect and crystal grain scale mechanics, such as dislocation accumulation under thermal loading \citep{Nascimento_etal_2025}, coupling heat conduction and gradient crystal plasticity \citep{Aldakheel_Miehe_2017},
the effect of heat conduction \citep{Li_etal_2019} and thermal boundary condition \citep{Connolly_etal_2020} on crystal plasticity model,
the thermal effect on transformation-induced plasticity \citep{Huang_etal_2025} and dynamic recrystallization \citep{Tao_etal_2019}, and application in the multiscale analysis \citep{Gierden_etal_2021,Schmidt_etal_2025} within the framework of thermal-crystal plasticity models.
The foundation for analytical models is gradually being established.
However, in most of these studies, the interpretation of simulation results has relied primarily on spatial snapshots at selected time instants or on average quantities over grains or regions, which makes it difficult to systematically extract the spatiotemporal structures governing the overall material response under cyclic thermal loading.

The difficulty in extracting the spatiotemporal structure of the mechanical response of polycrystals under cyclic thermal loading using the thermal-crystal plasticity analysis stems from the heterogeneous and complex stress-strain fields obtained from the crystal plasticity model.
This heterogeneity arises from the interactions among grains with different crystallographic orientations
\citep{Hansen_etal_2020,Thool_etal_2020,Pai_etal_2022,Pai_etal_2025},
and, in some cases, the coexistence of different crystal structures
\citep{Venkatraman_etal_2022,Shen_etal_2022}.
Additional sources of heterogeneity include microstructural evolution processes, such as the development of dislocation substructures \citep{Grilli_etal_2018,Dindarlou_Castelluccio_2022},
twinning or phase transformation \citep{Liu_etal_2023,Park_etal_2024},
recrystallization \citep{Min_etal_2020,Min_etal_2024},
hydrogen transport \citep{Park_etal2_2024,Park_etal_2025},
and damage accumulation
\citep{Loiodice_etal_2025,Sangid_etal_2025,Stopka_etal_2026}.
This inherent heterogeneity is a key advantage of crystal plasticity analysis,
enabling the reproduction of localized mechanical responses that cannot be captured by conventional plasticity models \citep{Montes_etal_2022,Aragon_etal_2024,Hu_etal_2024,Lim_etal_2025}.
However, it remains challenging to understand what characteristic responses are emerging for the material as a whole from the polycrystalline analysis results based on the crystal plasticity model.
Traditionally, post-processing methods have focused on either (i) evaluating the time-series changes in average values or representative points across the entire analysis model or specific regions, or (ii) visualizing the spatial distribution at a selected time point.
While other methods exist, such as focusing on specific grains or regions, most boil down to the approach described above.
However, the former approach risks losing information about spatial heterogeneity, while the latter makes it difficult to discern the temporal continuity.
As a result, it has remained unclear which spatiotemporal structures govern the macroscopic response of polycrystalline materials, particularly under cyclic thermal loading.

One method to overcome the shortage of conventional post-processing method is an analysis method based on dynamic mode decomposition (DMD) \citep{Schmid_2010,Kutz_etal_2016}.
DMD is widely used in the fluid dynamics field \citep{Asada_Kawai_2024,Takahashi_2025} and other fields \citep{Fujii_etal_2019,Filho_Santos_2019,Bruder_etal_2021,Dylewsky_etal_2022,Curtis_etal_2023} to extract the dominant spatiotemporal structure in the given dataset. 
It simultaneously determines dominant spatiotemporal modes and their evolution dynamics from time-series data, enabling low-dimensional representations that clarify underlying physical mechanisms.
Unlike proper orthogonal decomposition, which provides orthogonal spatial modes but does not explicitly capture their temporal dynamics, DMD simultaneously identifies both spatial structures and their characteristic time evolutions  \citep{Kutz_etal_2016}. 
Compared with direct machine-learning-based approaches, DMD maintains clear physical interpretability by representing system responses as superpositions of modal structures governed by linear dynamical behavior.
The temporal evolution of stress fields in polycrystalline materials is a phenomenon where periodic responses induced by thermal cycles coexist with localized responses, making systematic interpretation challenging. 
Applying DMD to the spatially heterogeneous and temporally evolving stress field allows the extraction of the primary modes describing the temporal evolution, enabling the mechanism to be organized based on both the spatial structure of the stress distribution and its temporal oscillation.
Furthermore, by leveraging DMD's reconstruction and prediction capabilities, response prediction based on the primary modes becomes possible without directly performing computationally expensive thermal-crystal plasticity analyses over extended periods.

The objective of this study is to establish an analysis framework that enables the systematic and low-dimensional extraction of dominant spatiotemporal structures from polycrystalline-scale thermal-crystal plasticity simulations.
Specifically, this study aims to analyze how thermal cycle frequency influences the temporal evolution of internal temperature and stress fields, leading to qualitatively different quasi-steady and unsteady response regimes. 
It is noted that such frequency-dependent behavior can be qualitatively anticipated from classical scaling arguments based on the Fourier number. 
Therefore, the role of the present simulations is not to establish this scaling law itself, but to examine how these thermally governed regimes manifest in spatially heterogeneous stress fields at the polycrystalline scale.

To this end, DMD is introduced as a diagnostic tool for thermal-crystal plasticity simulations. 
Rather than identifying new physical mechanisms, the objective is to provide a systematic framework to extract and quantify dominant spatiotemporal structures embedded in complex simulation data. 
By applying DMD to time-series stress-field data, this study aims to obtain a low-dimensional representation of the simulated fields and to characterize how their modal structure depends on thermal cycle frequency.

Through this framework, the present study provides a new methodology for organizing and interpreting high-dimensional thermal-mechanical simulation data, complementing conventional analyses based on averaged quantities or single snapshots.

\section{Methodology}
\label{Methodology}
\begin{figure*}[t]   
  \centering
  \includegraphics[width=\textwidth]{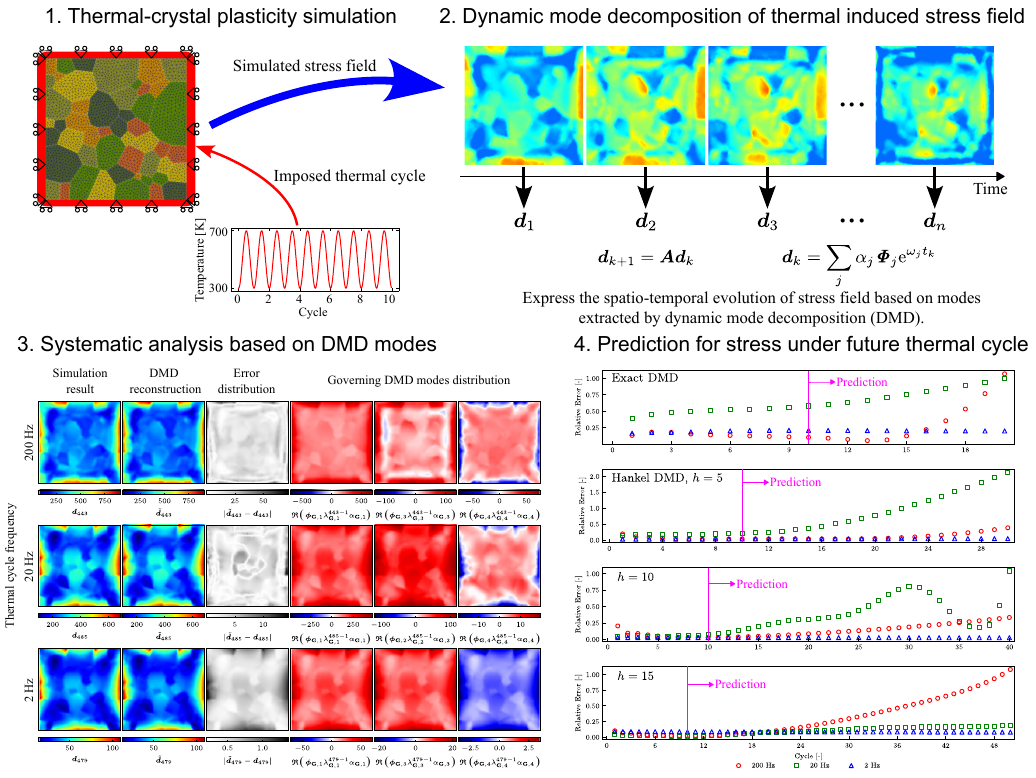}
  \caption{Schematic workflow of this study. The polycrystalline finite element model is first subjected to thermal cycle using coupled thermal-crystal plasticity analysis, producing time-series stress-field data. DMD is then applied to extract dominant spatiotemporal modes, enabling low-dimensional representation and prediction of thermal-mechanical responses.}
  \label{workflow}
\end{figure*}

In this study, the polycrystalline stress field under cyclic thermal loading at different thermal cycle frequencies is predicted using coupled thermal-crystal plasticity simulations.
DMD is then applied to the resulting time-series stress-field data to identify and analyze a reduced set of spatiotemporal modes that efficiently reconstruct the simulated stress-field fluctuations.
Finally, the feasibility of low-dimensional representation and prediction of thermal-mechanical responses based on the extracted modes is investigated.
Figure~\ref{workflow} provides an overview of the analysis framework, highlighting how high-dimensional stress-field data obtained from thermal-crystal plasticity simulations are transformed into a small number of physically interpretable spatiotemporal modes.

\subsection{Thermal-crystal plasticity coupling model}
\label{thermocpmodel}
In practical thermal-fatigue environments, such as those in additive manufacturing, the characteristic length scale of the temperature field is comparable to or even shorter than the polycrystalline microstructure \citep{Grilli_etal_2022,Pilgar_etal_2022,Nascimento_etal_2025}.
As a result, thermal stress evolves heterogeneously within and across grains, making polycrystalline-scale analysis essential for understanding the underlying thermal-mechanical response.
Accordingly, this study adopts a crystal plasticity model as the theoretical framework for describing the deformation field.

In the thermal-crystal plasticity analysis, the following two governing equations for the stress and temperature fields are solved in a coupled manner. 
The mechanical equilibrium equation, neglecting body force, for the Cauchy stress $\bm{T}$ is given by
\begin{equation}
  \mathrm{div} \bm{T}=\bm{0},
  \label{eqili}
\end{equation}
and the heat conduction equation for the temperature $\theta$ is expressed as 
\begin{equation}
    \rho c \frac{\partial \theta}{\partial t}=\nabla\cdot(k\nabla \theta)+\dot{q},
    \label{heatcond}
\end{equation}
where $\rho$ is the mass density, $c$ is the specific heat capacity, $k$ is the thermal conductivity, and $\dot{q}$ is the rate of internal heat generation.

When a material undergoes plastic deformation, much of the work is dissipated as heat.
To account for this thermal effect, the source term $\dot{q}$ on the right-hand side of Eq. \eqref{heatcond} is given by the following equation associated with the plastic work rate $\dot{W}^{\mathrm{p}}$ as in previous studies \citep{Roters_etal_2019,Li_etal_2019}:
\begin{equation}
  \dot{q}=\chi\dot{W}^{\mathrm{p}},
\end{equation}
where $\chi$ denotes the Taylor-Quinney parameter.

In this study, material deformation is described using a temperature-dependent crystal plasticity model.
First, the deformation gradient tensor $\bm{F}$ is decomposed into thermal expansion deformation $\bm{F}^{\mathrm{\theta}}$, plastic deformation $\bm{F}^{\mathrm{p}}$, and elastic deformation and rigid body rotation $\bm{F}^{\mathrm{*}}$ as follows 
\citep{Ozturk_etal_2016,Li_etal_2019}:
\begin{equation}
    \bm{F}=\bm{F}^{*}\bm{F}^{\mathrm{p}}\bm{F}^{\mathrm{\theta}}.
    \label{decomp}
\end{equation}
The multiplicative decomposition \eqref{decomp} introduces the first and second intermediate configurations in addition to the initial and current configurations.

The elastic constitutive equation is 
\begin{equation}
    \widetilde{\bm{T}}_{\mathrm{I\hspace{-1.2pt}I}}=\mathbb{C}^{\mathrm{e}}:\bm{E}^{\mathrm{e}}_{\mathrm{I\hspace{-1.2pt}I}}
    \label{elastic}
\end{equation}
where $\bm{E}^{\mathrm{e}}_{\mathrm{I\hspace{-1.2pt}I}}$ is the elastic Green-Lagrange strain, $\mathbb{C}^{\mathrm{e}}$ is the elastic coefficient, and $\widetilde{\bm{T}}_{\mathrm{I\hspace{-1.2pt}I}}$ is the pulled back Kirchhoff stress $\widetilde{\bm{T}}$ to the second intermediate configuration, defined respectively as follows:
\begin{equation}
    \widetilde{\bm{T}}_{\mathrm{I\hspace{-1.2pt}I}}=\bm{F}^{*-1}\widetilde{\bm{T}}\bm{F}^{*-\top}
    ,\quad
    \widetilde{\bm{T}}=J\bm{T}=J^{*}J^{\mathrm{\theta}}\bm{T} 
    \quad
    (\because J^{\mathrm{p}}=1).
\end{equation}
Here, $J$, $J^{*}$, $J^{\mathrm{p}}$, and $J^{\mathrm{\theta}}$ are the Jacobians of $\bm{F}$, $\bm{F}^{*}$, $\bm{F}^{\mathrm{p}}$, and $\bm{F}^{\mathrm{\theta}}$.

According to Eq. \eqref{decomp}, the velocity gradient tensor $\bm{L}=\dot{\bm{F}}\bm{F}^{-1}$ is decomposed as
\begin{equation}
    \bm{L}=\bm{L}^{*}+\bm{L}^{\mathrm{p}}+\bm{L}^{\mathrm{\theta}},
\end{equation}
where $\bm{L}^{*}$, $\bm{L}^{\mathrm{p}}$, and $\bm{L}^{\mathrm{\theta}}$ represent the velocity gradient due to the elastic deformation and rigid rotation, plastic deformation, and thermal expansion, respectively.

The velocity gradient due to thermal expansion $\bm{L}^{\mathrm{\theta}}$ is expressed as follows, assuming that the thermal expansion coefficient tensor is isotropic 
\citep{Ozturk_etal_2016,Li_etal_2019}
\begin{equation}
    \bm{L}^{\mathrm{\theta}}=\dot{\theta}\beta\bm{I},
\end{equation}
where $\beta$ is the thermal expansion coefficient.

The velocity gradient due to plastic deformation is expressed as follows, assuming that the slip deformation in each slip system:
\begin{equation}
    \bm{L}^{\mathrm{p}}=\sum_{\alpha}^{}
    \left(
        \bm{s}^{(\alpha)}\otimes\bm{m}^{(\alpha)}
    \right)\dot{\gamma}^{(\alpha)},
\end{equation}
where $\dot{\gamma}^{(\alpha)}$ is the slip rate of slip system $\alpha$, 
$\bm{s}^{(\alpha)}$ and $\bm{m}^{(\alpha)}$ are unit vectors in the slip direction and normal direction of the slip plane of the slip system $\alpha$.

Other possible plastic deformation mechanism in 316L stainless steel, such as deformation twinning, martensitic phase transformation, and creep, are not considered in the present model.
This simplification is justified because twinning and martensitic transformation can be neglected within the temperature range ($300$–$700,\si{\kelvin}$) and the relatively small accumulated strain considered in this study, as reported in the literature \citep{Molnar_2019,Sohrabi_2020,Maboudi_2026}. In addition, creep is negligible due to the short time scale of the thermal cycles (up to $5,\mathrm{s}$) \citep{Spigarelli_2026}.

To account for the effects of temperature changes, the slip rate $\dot{\gamma}^{(\alpha)}$ is described by the following slip rate hardening law that depends on temperature explicitly \citep{Patra_etal_2016}:
\begin{equation}
  \dot{\gamma}^{(\alpha)} =
  \begin{cases}
    0, & \text{if $\tau^{(\alpha)}_{\mathrm{eff}} \le 0$}, \\
    \dot{\gamma}_{0} \, \mathrm{sgn}(\tau^{(\alpha)})
    \exp \Biggl[ -\frac{\Delta F}{k_{\mathrm{b}}\theta} 
      \Bigl( 1 - \Bigl| \frac{\tau^{(\alpha)}_{\mathrm{eff}}}{g_{\mathrm{s}}} \Bigr|^p \Bigr)^q \Biggr], 
      & \text{if $\tau^{(\alpha)}_{\mathrm{eff}} > 0$},
  \end{cases}
  \label{sliphardeninglaw}
\end{equation}
where $\dot{\gamma}_{0}$ is the reference slip rate, $\tau^{(\alpha)}_{\mathrm{eff}}$ is the effective resolved shear stress, $\Delta F$ is the activation energy for dislocation, $k_{\mathrm{b}}$ is the Boltzmann constant, and $g_{\mathrm{s}}$ is the slip resistance due to the solid solution. 
Thermally activated slip-hardening laws such as Eq.~\eqref{sliphardeninglaw} adequately describe the slip deformation process in the region where the strain rate is $10^4\,\si{s}^{-1}$ or less \citep{Shahba_Ghosh_2016}. Since the slip rates actually observed in the analysis described below were at most on the order of $10\,\si{s}^{-1}$, they fall within the scope of Eq. \eqref{sliphardeninglaw}.

The effective resolved shear stress $\tau^{(\alpha)}_{\mathrm{eff}}$ is defined with the resolved shear stress $\tau^{(\alpha)}$ and passing resistance $g^{(\alpha)}_{\mathrm{p}}$ for glide dislocations due to the obstacles as 
\begin{equation}
    \tau^{(\alpha)}_{\mathrm{eff}}=\lvert\tau^{(\alpha)}\rvert-g^{(\alpha)}_{\mathrm{p}}.
\end{equation}

The passing resistance $g^{(\alpha)}_{\mathrm{p}}$ is expressed by Bailey-Hirsch equation 
to consider the interaction of glide dislocations with other dislocations as
\begin{equation}
    g^{(\alpha)}_{\mathrm{p}}=g_{\mathrm{p}0}+\alpha\mu b\sqrt{\rho^{(\alpha)}}
\end{equation}
with the initial slip resistance $g_{\mathrm{p}0}$, the Taylor hardening coefficient $\alpha$, the shear modulus $\mu$, the Burgers vector length $b$, and dislocation density $\rho^{(\alpha)}$.

Dislocation density $\rho^{(\alpha)}$ is decomposed into two types of dislocations: statistically stored dislocations (SSD) and geometrically necessary dislocations (GND), and is expressed as \citep{Salvini_etal_2024,Rezwan_etal_2025}
\begin{equation}
    \rho^{(\alpha)}
    =\sum_{\beta}^{}
    \omega^{(\alpha\beta)}
    \left(
        \rho^{(\beta)}_{\mathrm{SS}}
        +
        \lvert
            \rho^{(\beta)}_{\mathrm{e}}
        \rvert
        +
        \lvert
            \rho^{(\beta)}_{\mathrm{s}}
        \rvert
    \right),
\end{equation}
where $\omega^{(\alpha\beta)}$ is the dislocation interaction matrix, $\rho^{(\beta)}_{\mathrm{SS}}$ is the SSD density, and $\rho^{(\beta)}_{\mathrm{e}}$ and $\rho^{(\beta)}_{\mathrm{s}}$ are the edge and screw components of GND density.
SSD density $\rho^{(\alpha)}_{\mathrm{SS}}$ evolves with multiplication and annihilation as follows: 
\begin{equation}
    \dot{\rho}^{(\alpha)}_{\mathrm{SS}}=\left(k_{\mathrm{m}}\sqrt{\rho^{(\alpha)}}-2y_{\mathrm{c}}\rho^{(\alpha)}_{\mathrm{SS}}\right)\frac{1}{b}\lvert\dot{\gamma}^{(\alpha)}\rvert,
\end{equation}
where $k_{\mathrm{m}}$ is the accumulation rate and $y_{\mathrm{c}}$ is the critical distance for annihilation.
The edge and screw components of GND density are defined as 
\begin{equation}
    \dot{\rho}^{(\alpha)}_{\mathrm{e}}=-\frac{1}{b}\nabla\dot{\gamma}^{(\alpha)}\cdot\bm{s}^{(\alpha)},
\end{equation}
\begin{equation}
    \dot{\rho}^{(\alpha)}_{\mathrm{s}}=\frac{1}{b}\nabla\dot{\gamma}^{(\alpha)}\cdot\bm{t}^{(\alpha)},
\end{equation}
where $\bm{t}^{(\alpha)}=\bm{s}^{(\alpha)}\times\bm{m}^{(\alpha)}$.

\subsection{DMD framework}
\label{dmd}
This section describes the overview of DMD performed in this study \citep{Tu_etal_2014,Brunton_Kutz_2019}.
Detailed derivations of DMD framework are summarized in \ref{app1}.

Let $\bm{d}_{k}$ represent the data vector at the $k$-th time step, where $k=1,2,\dots,n$, and $n$ is the total number of time steps.
In the framework of DMD, the constant matrix $\bm{A}$, which satisfies the following relation, is assumed:
\begin{equation}
    \bm{D}_{2}\approx\bm{A}\bm{D}_{1},
    \label{dmd_start2}
\end{equation}
where $\bm{D}_{1}=[\bm{d}_{1},\bm{d}_{2},...,\bm{d}_{n-1}]$ and $\bm{D}_{2}=[\bm{d}_{2},\bm{d}_{3},...,\bm{d}_{n}]$.
DMD provides an approximation of the eigenvalues and eigenvectors of the time-evolution operator $\bm A$ from the snapshot matrices $\bm{D}_{1}$ and $\bm{D}_{2}$. 
The following eigenvalue decomposition for $\widetilde{\bm{A}}$ obtains eigenvalues of $\bm{A}$:
\begin{equation}
    \widetilde{\bm{A}}=\bm{V}\bm{\mathit{\Lambda}}\bm{V}^{-1}.
\end{equation}
Here, $\widetilde{\bm{A}}$ is constructed using the singular value decomposition of $\bm{D}_{1}$, following the standard DMD formulation \citep{Brunton_Kutz_2019}.

In the framework of Exact DMD \citep{Tu_etal_2014}, the eigenvectors of $\bm{A}$ are defined as follows:
\begin{equation}
    \bm{\mathit{\Phi}}=\bm{D}_{1}\widetilde{\bm{W}}\widetilde{\bm{\mathit{\Sigma}}}^{-1}\bm{V}\bm{\mathit{\Lambda}}^{-1},
    \label{exactvector2}
\end{equation}
where the matrix $\bm{\mathit{\Phi}}$ consists of the eigenvectors of $\bm{A}$, and $\widetilde{\bm{W}}$ and $\widetilde{\bm{\mathit{\Sigma}}}$ are obtained from the singular value decomposition of $\bm{D}_{1}$.

Reconstruction of the original data $\bm{d}_{k}$ using DMD mode can be performed by
\begin{equation}
    \bm{d}_{k}
    =
    \bm{\mathit{\Phi}}\bm{\mathit{\Lambda}}^{k-1}\bm{\mathit{\Phi}}^{\dagger}\bm{d}_{1}
    =\bm{\mathit{\Phi}}\bm{\mathit{\Lambda}}^{k-1}\bm{\alpha}
    =\sum_{j}
    \bm{\phi}_{j}\lambda^{k-1}_{j}\alpha_{j},
    \label{reconstruction}
\end{equation}
where the vector $\bm{\phi}_{j}$ represents the $j$-th column vector of the matrix $\bm{\mathit{\Phi}}$ and $\lambda_{j}$ represents the diagonal value of the $j$-th column vector of the matrix $\bm{\mathit{\Lambda}}$. $^{\dagger}$ represents the Moore-Penrose pseudo inverse.
The coefficient of each DMD mode $\bm{\alpha}$ is defined by
\begin{equation}
    \bm{\alpha}=\bm{\mathit{\Phi}}^{\dagger}\bm{d}_{1}.
    \label{alphavec}
\end{equation}

Furthermore, the amplitude $\sigma_{j}$ and frequency $f_{j}$ of the $j$-th DMD mode, obtained by converting the discrete eigenvalue $\lambda_{j}$ to the continuous eigenvalue $\omega_{j}$, are given by the following expressions when the time interval between each time-series data point is $\Delta t$ \citep{Asada_Kawai_2024}:
\begin{equation}
    \sigma_{j}=\frac{\mathfrak{R} \{\log(\lambda_{j})\}}{\Delta t}
    ,\quad
    f_{j}=\frac{\mathfrak{I} \{\log(\lambda_{j})\}}{2\pi\Delta t},
    \label{freqdef}
\end{equation}
where $\mathfrak{R}(\circ)$ and $\mathfrak{I}(\circ)$ represent the real and imaginary part of $\circ$.

In the Exact DMD formulation described above, the data vector $\bm{d}_{k}$ represents the system state at a single time step $k$.
In contrast, Hankel DMD extends the state vector by incorporating time-delayed snapshots, thereby enriching the information used to approximate the linear time-evolution operator.
Hankel DMD has been widely used to enhance data reconstruction capabilities for low-dimensional data \citep{Fujii_etal_2019,Filho_Santos_2019}, and recent studies have reported its effectiveness for high-dimensional data as well \citep{Asada_Kawai_2024}.

Specifically, for a given time-delay embedding dimension $h$, we define the extended data vector $\bm{h}^{(h)}_{k}$ as
\begin{equation}
    \bm{h}^{(h)}_{k}=
    \begin{bmatrix}
        \bm{d}_{k}\\
        \bm{d}_{k+1}\\
        \vdots\\
        \bm{d}_{k+h-1}
    \end{bmatrix}.
    \label{extendeddatavector}
\end{equation}
By using the extended data vector \eqref{extendeddatavector}, the Hankel matrix is introduced as follows: 
\begin{equation}
    \bm{H}^{(h)}
    =
    \begin{bmatrix}
        \bm{h}^{(h)}_{1}\cdots\bm{h}^{(h)}_{n-h+1}
    \end{bmatrix}.
\end{equation}
Hankel DMD uses the following data matrix $\bm{H}^{(h)}_{1}$ and $\bm{H}^{(h)}_{2}$ instead of $\bm{D}_{1}$ and $\bm{D}_{2}$:
\begin{equation}
    \bm{H}^{(h)}_{1}=
    \begin{bmatrix}
        \bm{h}^{(h)}_{1}\cdots\bm{h}^{(h)}_{n-h},
    \end{bmatrix}
\end{equation}
\begin{equation}
    \bm{H}^{(h)}_{2}=
    \begin{bmatrix}
        \bm{h}^{(h)}_{2}\cdots\bm{h}^{(h)}_{n-h+1}.
    \end{bmatrix}
\end{equation}
Equation~\eqref{dmd_start2} is rewritten as follows:
\begin{equation}
    \bm{H}^{(h)}_{2}\approx\bm{A}\bm{H}^{(h)}_{1}.
    \label{hankel_dmd_start_2}
\end{equation}
The procedures to obtain the eigen modes of matrix $\bm{A}$ in Eq.~\eqref{hankel_dmd_start_2} are the same as the case of Exact DMD.
Under the notation above, it is clear that the Exact DMD corresponds to the case of $h=1$ for the Hankel DMD.

\subsection{Numerical implementation and analysis conditions}
\subsubsection{Thermal-crystal plasticity simulation}
\label{thermal_cp_condition}
In the thermal-crystal plasticity simulation, the governing equations \eqref{eqili} and \eqref{heatcond} are solved simultaneously using a monolithic solver.
The finite element analysis is performed using the Multiphysics Object Oriented Simulation Environment (MOOSE) framework \citep{Giudicelli_etal_2024}.

The material studied in this research is 316L stainless steel, which is widely used in applications subjected to thermal cycle and is also commonly employed in additive manufacturing.
The parameters used are shown in Table~\ref{parameters}.
Literature values \citep{Kim_1975,Scherer_etal_2024} are used for material constants.
Although the thermal conductivity and specific heat of 316L stainless steel exhibit temperature dependence in the temperature range between $300\,\si{\kelvin}$ and $700\,\si{\kelvin}$ \citep{Kim_1975}, they are treated as constants in this study.

This simplification is introduced to isolate the effect of thermal cycle frequency from additional non-linearities associated with material properties.
The influence of temperature-dependent thermal properties is quantitatively assessed in \ref{kc_temp}. 
Within the present parameter range, their impact on the simulation results is found to be smaller than that induced by variations in thermal cycle frequency. 
The additional simulations incorporating temperature-dependent properties show consistent trends with the present results, supporting the validity of this modeling assumption for the purpose of the study.

In addition, the effect of anisotropic grain boundary thermal resistance is not explicitly considered. 
To provide a rough estimate of its potential impact, a conservative upper-bound calculation can be performed. 
The bulk thermal resistance within a grain $R_{\mathrm{grain}}$ is approximated as 
$R_{\mathrm{grain}} = L_{\mathrm{grain}} / k$,
where the grain size $L_{\mathrm{grain}} = 100\,\si{\micro\metre}$ and the thermal conductivity $k = 13.96\,\si{\watt\metre^{-1}\kelvin^{-1}}$, giving 
$R_{\mathrm{grain}} \approx 7.16 \times 10^{-6}\,\si{\metre^{2}\kelvin\watt^{-1}}$.
Even if a relatively large grain-boundary thermal resistance $R_{\mathrm{gb}}$ on the order of $R_{\mathrm{gb}} \sim 10^{-8}\,\si{\metre^{2}\kelvin\watt^{-1}}$
is assumed from the value of ceramic \citep{Smith_etal_2018}, the ratio $R_{\mathrm{gb}}/R_{\mathrm{grain}}$ remains on the order of $10^{-3}$. 
This suggests that the contribution of grain boundaries is negligible compared to bulk conduction at the present polycrystalline length scale.

This conclusion is further supported by experimental studies on FCC-structured TWIP steels with grain sizes of order $100\,\si{\micro\metre}$, which show that bulk thermophysical properties such as thermal conductivity and thermal expansion are largely insensitive to grain size \citep{Hwang_2025}. 
Although direct measurements of grain-boundary thermal resistance in 316L stainless steel are not available, the minimal grain-size effect on bulk transport implies that the temperature drop across individual grain boundaries is expected to be small relative to the imposed thermal gradients. 
Therefore, the assumption of isotropic and spatially uniform thermal properties, effectively neglecting grain-boundary resistance, is considered reasonable for the present simulations.

The Taylor-Quinney parameter may depend on strain, strain rate, and microstructure \citep{Soares_Hokka_2021}. However, as confirmed in \ref{sensitivity_chi}, the influence of this parameter’s value on the analysis results is negligible in the thermal cycle analysis conducted in this study.

\begin{sidewaystable*}[htbp]
\centering\caption{Model parameters used in thermal-crystal plasticity simulation}\label{parameters}
\begin{tabular}{ccc}\hline
    Mechanical property \citep{Scherer_etal_2024} & Value & Unit\\
    \hline
    Component of elastic coefficient tensor, $C_{11}$ & $233360-51.3\theta(\mathrm{K})$ &$\si{\mega\pascal}$\\
    Component of elastic coefficient tensor, $C_{12}$ & $152880-27.7\theta(\mathrm{K})$ &$\si{\mega\pascal}$\\
    Component of elastic coefficient tensor, $C_{44}$ & $134430-30.8\theta(\mathrm{K})$ &$\si{\mega\pascal}$\\
    Shear modulus, $\mu$ & $0.02478\theta (\mathrm{K})^2-56.01\theta (\mathrm{K})+86070$ & $\si{\mega\pascal}$\\ 
    Burgers vector length, $b$ & 0.254 &$\si{\nano\metre}$\\
    \hline
    Thermal property \citep{Kim_1975} & Value & Unit \\ 
    \hline
    Mass density, $\rho$ & $7954$ & $\si{\kilo\gram\,\metre^{-3}}$\\
    Thermal conductivity, $k$ & $13.96$ & $\si{\watt\metre^{-1}\kelvin^{-1}}$ \\
    Specific heat capacity, $c$ & $498.7328$ & $\si{\joule\,\kilo\gram^{-1}\kelvin^{-1}}$\\
    Thermal expansion coefficient, $\beta$ & $1.86357328\times 10^{-5}$ & $\si{\kelvin}^{-1}$ \\
    \hline
    Model parameter & Value & Unit \\
    \hline
    Taylor-Quinney parameter, $\chi$ & $0.85$ & $-$\\
    Reference slip rate, $\dot{\gamma}_{0}$ & $10^{4}$ &$\mathrm{s}^{-1}$\\
    Activation energy for dislocation, $\Delta F$ & $6.63\times 10^{-19}$ &$\si{\joule}$\\
    Slip resistance due to the solid solution, $g_{\mathrm{s}}$ & $130$ & $\si{\mega\pascal}$\\
    Initial slip resistance, $g_{\mathrm{p}0}$ & $20$ &$\si{\mega\pascal}$\\
    Taylor hardening coefficient, $\alpha$ & $0.2$ & $-$\\
    Self hardening magnitude of $\omega^{(\alpha\beta)}$ &1 & $-$\\
    Latent hardening magnitude of $\omega^{(\alpha\beta)}$ &1.4&$-$\\
    Initial statistical stored dislocation density, $\rho_{\mathrm{SS}}^{(\alpha)}$& $4\times 10^{9}$ & $\si{\metre}^{-2}$ \\
    Dislocation accumulation rate, $k_{\mathrm{m}}$ & $0.08$ & $-$ \\
    Critical annihilation distance for adjacent dislocations, $y_{\mathrm{c}}$ & $0.6$ & $\si{\nano\metre}$ \\
    Parameter controlling the glide resistance profile, $p$ & $0.75$ & $-$ \\
    Parameter controlling the glide resistance profile, $q$ & $1.33333333$ & $-$ \\ 
    \hline
\end{tabular}
\end{sidewaystable*}

\begin{figure}[t]
\begin{center} 
\includegraphics[width=0.45\textwidth]{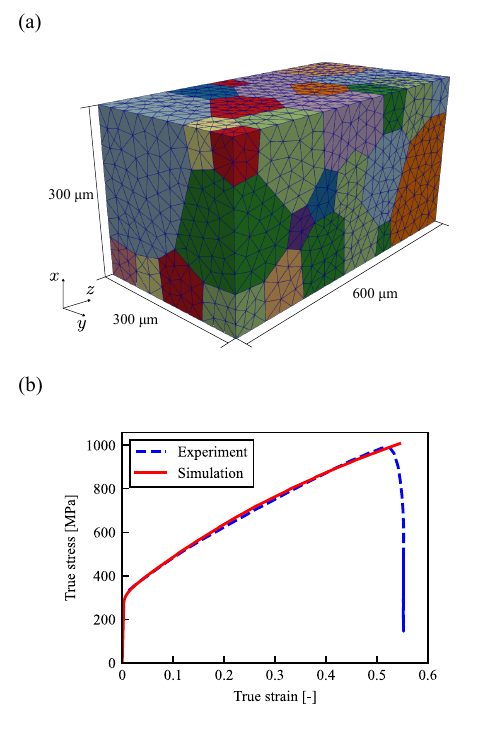}
\caption{Calibration of crystal plasticity model parameters. (a) 54 grains finite element model used for calibration. (b) Comparison of simulated and experimental stress-strain curves under tensile loading.} 
\label{cp_calib} 
\end{center}
\end{figure}

To calibrate some model parameters related to the dislocation density evolution, we performed an analysis simulating the uniaxial tensile test on 316L stainless steel reported by a previous study \citep{Yan_etal_2012}.
The polycrystalline model used in the analysis is shown in Fig.~\ref{cp_calib}(a).
The dimensions of this model with 54 grains are $0.3\,\si{\milli\metre}\times 0.3\,\si{\milli\metre}\times 0.6\,\si{\milli\metre}$. 
The computational model is created using 14,904 quadratic tetrahedral elements with the grain growth model by Neper \citep{Quey_etal_2011,Quey_etal_2018}, ensuring the average grain size is consistent with the reported $100\,\si{\micro\metre}$ \citep{Yan_etal_2012}.
The boundary conditions are specified as follows:
the lateral surfaces of the specimen are free, the bottom surfaces is fixed in the $z$-direction, while to prevent rigid body motion, all degrees of freedom of the node at the origin are constrained, and the top surface is subjected to a prescribed displacement in the $z$-direction corresponding to a constant strain rate.
The applied strain rate is set to $5\times 10^{-3}\,\si{s^{-1}}$, consistent with the experimental conditions.
A comparison of the reported stress-strain curve with the analytical results obtained using the parameters in Table~\ref{parameters} is shown in Fig.~\ref{cp_calib}(b).

Note that the crystal plasticity model used in this study accounts for the effects of temperature in the strain rate hardening law and the elastic modulus. 
We verified whether this model adequately accounts for the effects of temperature-induced softening in \ref{thermal_softening}.

The three-dimensional tensile analysis is used solely for calibration of crystal plasticity parameters to reproduce the basic mechanical response.
Since the present crystal plasticity model does not include parameters specifically requiring calibration under thermal cycling conditions, no direct comparison with thermal-cycle experiments is performed. 
In contrast, the thermal cycle simulations described below aim to investigate the evolution and frequency dependence of stress fields driven by imposed temperature gradients, representing a different physical setting.

Thermal cycle analysis is performed on an analytical model expressing the polycrystalline structure shown in Fig.~\ref{workflow}(a). 
The polycrystalline structure, consisting of 50 grains, was also created using Neper's grain growth model \citep{Quey_etal_2011,Quey_etal_2018} as before, with dimensions of $0.7\,\si{\milli\metre}\times 0.7\,\si{\milli\metre}$ and employing second-order triangular elements. 
The four edges of the polycrystalline structure were fixed by constraints as shown in Fig.~\ref{workflow}(a).
To reduce computational cost while resolving the evolution of grain-scale stress fields under thermal cycling, a two-dimensional plane-strain model is adopted.

The plane-strain assumption introduces an out-of-plane constraint, which may alter the stress state and consequently the extent of plastic deformation.
Such changes could, in principle, affect the temperature field through heat generation by plastic work.
However, as demonstrated in \ref{sensitivity_chi}, additional simulations with the Taylor–Quinney coefficient set to zero show nearly identical temperature and stress evolution compared to the reference case.
This indicates that the contribution of heat dissipation due to the plastic deformation to the temperature field is limited under the present conditions.

Therefore, the temperature field is governed primarily by the imposed thermal boundary conditions and heat conduction and is insensitive to variations in the mechanical response.
While the plane-strain approximation affects the absolute stress state and plastic strain magnitude, the present study focuses on relative frequency-dependent trends in stress heterogeneity rather than quantitative stress levels.

The temperature on these four edges is denoted $\theta_{\mathrm{bc}}$ and controlled by a sinusoidal wave ranging from $\theta_{\mathrm{low}}=300\,\si{\kelvin}$ to $\theta_{\mathrm{high}}=700\,\si{\kelvin}$ as follows:
\begin{equation}
    \theta_{\mathrm{bc}}
    =
    \frac{\theta_{\mathrm{low}}+\theta_{\mathrm{high}}}{2}
    +
    \frac{\theta_{\mathrm{high}}-\theta_{\mathrm{low}}}{2}
    \sin\left(
        2\pi f t - \frac{\pi}{2}
    \right).
    \label{thetabc}
\end{equation}
Such thermal loading can be regarded as a simplified representation of the repeated heating and cooling encountered in practical applications, such as cyclic thermal loading during operation or manufacturing processes.
The frequency of the sinusoidal wave $f$ is set to $2\,\mathrm{Hz}$, $20\,\mathrm{Hz}$, and $200\,\mathrm{Hz}$.
These frequencies are selected based on the temperature changing rate observed or assumed in previous studies on additive manufacturing processes \citep{Upadhyay_etal_2021,Nascimento_etal_2025}. 
The time step width is fixed at $10^{-4}\,\si{s}$, $10^{-3}\,\si{s}$, and $10^{-2}\,\si{s}$ for thermal cycle frequencies of $200\,\mathrm{Hz}$, $20\,\mathrm{Hz}$, and $2\,\mathrm{Hz}$ to guarantee sufficient time resolution and an efficient computational cost.

Under the present thermal cycling conditions, the loading path differs from conventional mechanical cyclic loading involving full tension-compression reversals, as the stress evolution is primarily driven by spatially non-uniform thermal expansion under constrained condition, and the imposed temperature does not decrease below the initial value.
Consequently, the deformation conditions considered here are distinct from those under which cyclic plasticity phenomena such as the Bauschinger effect are typically pronounced.

Furthermore, this thermal cycle analysis does not account for the effects of defects such as voids and precipitates.
However, these defects are commonly found in materials subjected to thermal cycling and are believed to have a significant impact on the stress response under thermal loading. 
Therefore, we conducted an analyses that take the distributed voids and precipitates into account, and the results are summarized in \ref{void_particle}.

Although Eq.~\eqref{thetabc} imposes a strictly periodic thermal boundary condition in this study, the subsequent DMD and Hankel DMD analyses do not require the input to be exactly periodic.
DMD have been successfully applied to non-periodic or noisy datasets in various fields such as \cite{Brunton_etal_2016,Takahashi_2025}.
In \ref{appf}, we investigate the DMD’s reconstruction ability in response to non-periodic heat input.
For more general nonlinear or fully non-periodic thermal inputs, Hankel DMD can be interpreted within the Koopman operator framework, and increasing the embedding dimension $h$ approaches a Koopman representation, potentially allowing prediction under such conditions \citep{Arbabi_etal_2017,Bruder_etal_2021}.

\subsubsection{DMD condition}
In this study, DMD is applied to the temporal evolution of the von Mises stress field obtained from the thermal-crystal plasticity simulations.
The von Mises stress is selected as the target quantity because plastic deformation and stress evolution under thermal cycle are of primary interest.
As the von Mises stress serves as a representative measure of yielding and accumulated plastic response at the macroscopic level, it provides a suitable scalar quantity for assessing the influence of thermal cycle frequency on plastic-dominated stress evolution.
Although the computational cost increases, DMD itself can be applied to 3D data \citep{Ohmichi_2017,Asada_Kawai_2024,Takahashi_2025}, and the hyperparameter selection strategy described below remains essentially the same. 
From a computational standpoint, the difference between 2D and 3D data is reflected in the dimension $n$ of the data vector, which becomes $n = n_x \times n_y$ for 2D data and $n = n_x \times n_y \times n_z$ for 3D data, where $n_x$, $n_y$, and $n_z$ denote the number of grid points in each spatial direction. 
Since the dominant operations in DMD, such as singular value decomposition, scale with the size of the data matrix, the computational cost increases proportionally with the number of spatial discretisation points. 
Therefore, the computational efficiency of DMD for 3D problems can be estimated directly based on the increase in data size.

To apply DMD consistently across different simulations, the stress field at each time step is interpolated onto a uniform Cartesian grid of $350 \times 350$ points using the \texttt{sample} function implemented in PyVista \citep{Sullivan_2019}.
The interpolated stress field at the $k$-th time step is then flattened into a column vector, which is used as the data vector $\bm{d}_{k}$ defined in Section~\ref{dmd}.
The thermal-crystal plasticity simulations are performed from $t=0$ for ten thermal cycles for each thermal cycle frequency.
Snapshots of the stress field are output every ten time steps, resulting in 501 snapshots for each simulation.
However, the stress field at $t=0$ is identically zero due to the absence of thermal loading and therefore does not contain meaningful dynamical information.
To avoid introducing a trivial zero vector into the DMD analysis, the snapshot at the tenth time step is defined as the initial data vector $\bm{d}_{1}$.
Consequently, the total number of snapshots used for DMD analysis is 492.

The time-delay embedding dimension $h$ determines the extent to which temporal correlations are taken into account.
Compared to Exact DMD, Hankel DMD increases the row dimension of the matrix $\bm{A}$ from $n$ to $nh$, while the column dimension is reduced by a factor of $h$, where $n$ denotes the dimension of the original data vector $\bm{d}_{k}$.
In general, $h$ should be selected based on the characteristic time scale of the system so that the dominant temporal correlations are sufficiently captured.
At the same time, excessively large values of $h$ increase the computational cost and may degrade numerical stability.
In this study, the maximum value of $h$ is set to 15, considering that each thermal cycle consists of 50 time steps and that $h$ should capture a sufficient portion of the temporal correlation while avoiding high computational cost.

The number of modes used to represent the system is determined based on a trade-off between reconstruction accuracy and model reduction.
In practice, only a limited number of modes contribute significantly to the reconstruction of the major spatiotemporal variations in the dataset, while the remaining modes mainly represent noise or weak fluctuations.
Therefore, a mode selection procedure is required to identify the physically relevant modes systematically. 
In this study, a greedy mode selection strategy widely used in previous studies \citep{Ohmichi_2017,Asada_Kawai_2024,Takahashi_2025} is employed.
A mode selection method utilizing L1 regularization has also been proposed \citep{Jovanovic_etal_2014}, and both methods are reported effective in \cite{Ohmichi_2017}.

For Hankel DMD with a time-delay embedding dimension $h$, the reconstruction error of the Hankel matrix using $m$ selected modes is defined as
\begin{equation}
    \epsilon^{(h,m)}=\frac{\Vert\bm{H}^{(h)}-\widetilde{\bm{H}}^{(h,m)}\Vert_{\mathrm{F}}}{\Vert \bm{H}^{(h)} \Vert_{\mathrm{F}}}
    \label{recon_error_def}
\end{equation}
where $\bm{H}^{(h)}$ denotes the original Hankel matrix constructed from the data sequence and $\widetilde{\bm{H}}^{(h,m)}$ represents the reconstructed Hankel matrix using $m$ selected DMD modes, defined as
\begin{equation}
    \widetilde{\bm{H}}^{(h,m)}=
    \begin{bmatrix}
        \tilde{\bm{h}}^{(h,m)}_{1}\cdots\tilde{\bm{h}}^{(h,m)}_{n-h}    
    \end{bmatrix}.
\end{equation}
Each reconstructed column vector $\tilde{\bm{h}}^{(h,m)}_{k}$ is obtained by
\begin{equation}
    \tilde{\bm{h}}^{(h,m)}_{k}
    =\sum_{j=1}^{m}
    \bm{\phi}_{\mathrm{G},j}\lambda^{k-1}_{\mathrm{G},j}\alpha_{\mathrm{G},j}
    \label{reconstruction2}
\end{equation}
where the subscript ``G'' indicates that the DMD modes, eigenvalues, and corresponding modal coefficients are ordered according to their contribution to reducing the reconstruction error.
In the present analysis, the maximum number of retained modes is fixed to $m=20$ based on the following results.

\section{Results and discussion}
\label{ResultsAndDiscussion}
\subsection{Thermal-mechanical response under different thermal cycle frequencies}
\label{ResultsAndDiscussion1}
\subsubsection{Time history of spatially averaged temperature and spatial distribution of temperature at the end of 10th cycle}
\begin{figure*}[t]   
  \centering
  \includegraphics[width=\textwidth]{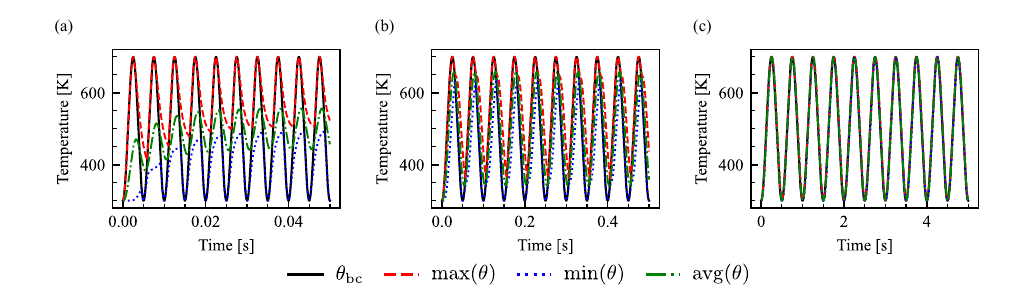}
  \caption{Time evolution of temperature for (a)$200\,\mathrm{Hz}$, (b)$20\,\mathrm{Hz}$, and (c)$2\,\mathrm{Hz}$. Black, red, blue, and green lines indicate the boundary, maximum, minimum, and spatially averaged temperatures, respectively.}
  \label{temperature_time_history}
\end{figure*}

\begin{figure*}[t]   
  \centering
  \includegraphics[width=\textwidth]{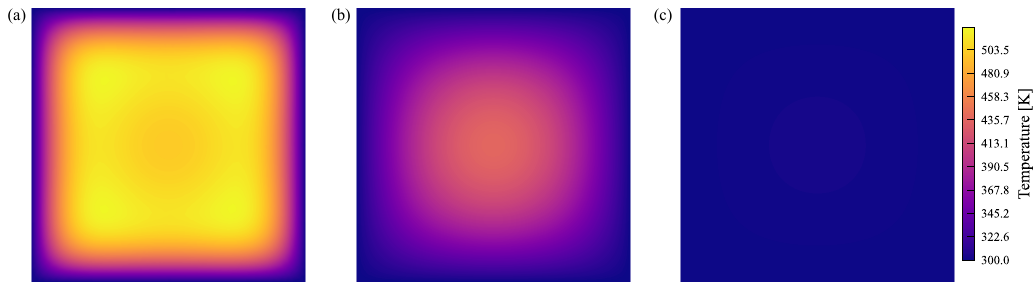}
  \caption{Temperature distributions obtained from thermal cycle analyses at the end of the 10th cycle for (a)$200\,\mathrm{Hz}$, (b)$20\,\mathrm{Hz}$, and (c)$2\,\mathrm{Hz}$.}
  \label{temp_heatmap_10cycle}
\end{figure*}

Figure~\ref{temperature_time_history} compares the time histories of the boundary, maximum, minimum, and spatially averaged temperatures obtained from the thermal-crystal plasticity simulations.
Among the three cases, the $200\,\mathrm{Hz}$ case exhibited the largest temperature difference at the same time point, followed by the $20\,\mathrm{Hz}$ case.
On the other hand, for the $2\,\mathrm{Hz}$ case, the maximum, minimum, and average temperatures oscillated in near-perfect agreement throughout all time intervals.
Furthermore, at $20\,\mathrm{Hz}$, the waveforms of the maximum, minimum, and average temperatures were nearly identical from cycle to cycle.
In contrast, at $200\,\mathrm{Hz}$, the temperature oscillation waveforms exhibited subtle but systematic changes as the cycles progressed.
Specifically, the lower envelope of the maximum temperature within each cycle gradually increased, while the overall amplitude of oscillations in minimum temperatures gradually increased.

The spatial distributions of temperature for each case at the end of the 10th cycle are shown in Fig.~\ref{temp_heatmap_10cycle} .
The $200\,\mathrm{Hz}$ case exhibited the largest temperature difference within the analysis domain, followed by the $20\,\mathrm{Hz}$ case.
In this study, since isotropic thermal conductivity properties were considered, the influence of the polycrystalline microstructure shape does not strongly manifest in the temperature field itself.
In the $200\,\mathrm{Hz}$ case, the peak temperatures are distributed away from the domain center, whereas in the $20\,\mathrm{Hz}$ case, the peak temperature is located near the center.
In contrast, the temperature distribution for the $2\,\mathrm{Hz}$ case was nearly uniform, with almost no spatial variation.
These characteristics are consistent with the temperature time histories discussed above.

The temporal evolution of temperatures differs at different frequencies because the Fourier number, a non-dimensional number that characterizes the relative importance of thermal diffusion over a given time and length scale, varies with frequency.
The Fourier number is defined as 
\begin{equation}
    \mathrm{Fo} = \frac{k\,t_{\mathrm{c}}}{\rho c \,L_{\mathrm{c}}^{2}},
\end{equation}
where $t_{\mathrm{c}}$ is a characteristic time scale, and $L_{\mathrm{c}}$ is a characteristic length scale of the system.
In the present study, the characteristic time scale $t_{\mathrm{c}}$ is taken as the time period of the cyclic temperature input, $t_{\mathrm{c}}=1/f$, and the characteristic length scale $L_{\mathrm{c}}$ corresponds to a representative dimension of the analysis domain.
For high-frequency cases such as $200\,\mathrm{Hz}$ and $20 \,\mathrm{Hz}$, the resulting Fourier numbers are about $0.025$ and $0.25$, indicating that thermal diffusion is not sufficiently fast to accommodate the rapid boundary temperature changes.
As a consequence, heat cannot propagate into the interior within a single cycle, leading to pronounced spatial temperature inhomogeneities, as observed in the simulations.
In contrast, for the $2\,\mathrm{Hz}$ case, the larger Fourier number, about $2.5$, implies that thermal diffusion is fast relative to the boundary temperature changes.
As a result, the internal temperature closely follows the boundary temperature, yielding an almost spatially uniform temperature field.
It should be emphasized that the absolute values of the applied frequencies do not have intrinsic physical meaning by themselves.
Their influence on the thermal response must always be interpreted in relation to the time and spatial scales of the system and its thermal transport properties using non-dimensional parameters such as the Fourier number $\mathrm{Fo}$.

\subsubsection{Time history of spatially averaged von Mises stress and equivalent plastic strain}
\begin{figure*}[t]   
  \centering
  \includegraphics[width=\textwidth]{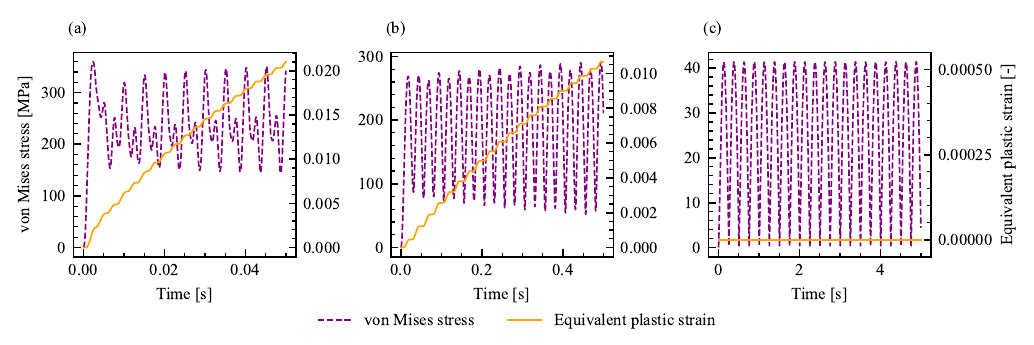}
  \caption{Time evolution of von Mises stress and equivalent plastic strain for (a)$200\,\mathrm{Hz}$, (b)$20\,\mathrm{Hz}$, and (c)$2\,\mathrm{Hz}$.}
  \label{vm_eqps_time_history}
\end{figure*}

Figure~\ref{vm_eqps_time_history} compares the time histories of the spatially averaged von Mises stress at $200\,\mathrm{Hz}$, $20\,\mathrm{Hz}$, and $2\,\mathrm{Hz}$.
The magnitude of spatially averaged von Mises stress obtained from our thermal-crystal plasticity simulations is comparable to previously reported computational and experimental values for 316L stainless steel under similar thermal loading conditions \citep{Grilli_etal_2022,Simson_2017,Santa-aho_2021}.

At $200\,\mathrm{Hz}$, the waveform in the initial cycle differs from those in the subsequent cycles, and the maximum peak stress attained in the initial cycle is never exceeded thereafter.
from the third cycle onward at $200\,\mathrm{Hz}$, as well as throughout the $20\,\mathrm{Hz}$ case, two local maxima appear within a single cycle, with the larger separation between two peaks observed at $200\,\mathrm{Hz}$.
Although two local maxima are also present at $2\,\mathrm{Hz}$, their separation is much smaller than in the higher-frequency cases.
Furthermore, in both the $200\,\mathrm{Hz}$ case (from the third cycle onward) and the $20\,\mathrm{Hz}$ case, the stress amplitude increases sharply with increasing cycle number.
The corresponding time history of the spatially averaged equivalent plastic strain (Fig.~\ref{vm_eqps_time_history}) indicates that the largest plastic strain accumulation occurred at $200\,\mathrm{Hz}$, while that at $20\,\mathrm{Hz}$ is approximately half as large. 
On the other hand, plastic strain accumulation was negligible at $2\,\mathrm{Hz}$.

In the present analysis, the primary source of stress generation is thermal expansion induced by temperature variations, combined with surrounding deformation constraints, which provides insight into the results shown in Fig.~\ref{vm_eqps_time_history} discussed above.
When the resulting thermal stress exceeds the yield stress, plastic deformation occurs, and plastic strain accumulates.
The difference between the stress waveform in the initial cycle at $200\,\mathrm{Hz}$ shown in Fig.~\ref{vm_eqps_time_history} can be attributed mainly to the non-stationary temperature history. 
As shown in Fig.~\ref{temperature_time_history}, the minimum temperature profile during the initial cycle at $200\,\mathrm{Hz}$ does not coincide with those in the subsequent cycles, indicating that the temperature fields have not yet reached a periodic steady state.
The non-stationarity in the minimum temperature profile is reflected in the temporal evolution of thermal expansion and, consequently, in the thermal stress history, resulting in a distinct stress waveform in the initial cycle.
For the $2\,\mathrm{Hz}$ case, which exhibited the smallest stress oscillation range in the spatially averaged time history, the thermal stress did not exceed the yield stress, resulting in negligible plastic strain accumulation.

\subsubsection{Spatial distributions of von Mises stress and equivalent plastic strain at the end of the 10th cycle}

\begin{figure*}[t]   
  \centering
  \includegraphics[width=\textwidth]{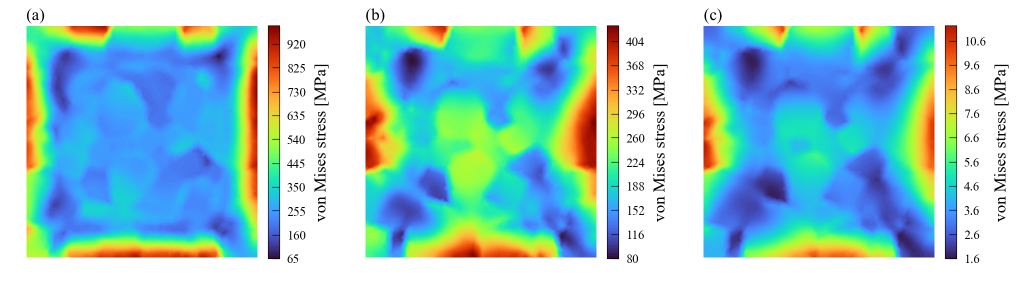}
  \caption{von Mises stress distributions obtained from thermal cycle analyses at the end of the 10th cycle for (a)$200\,\mathrm{Hz}$, (b)$20\,\mathrm{Hz}$, and (c)$2\,\mathrm{Hz}$.}
  \label{vm_heatmap_10cycle}
\end{figure*}
\begin{figure*}[t]   
  \centering
  \includegraphics[width=\textwidth]{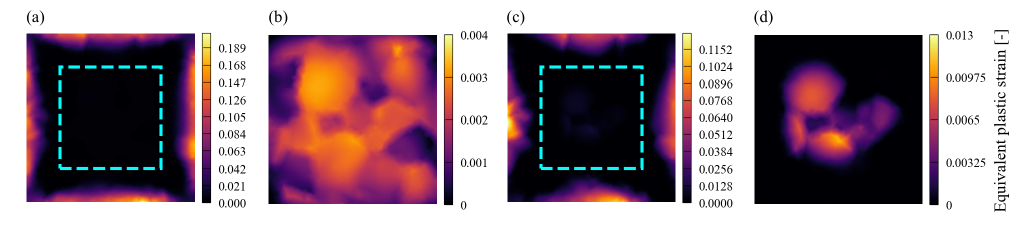}
  \caption{Equivalent plastic strain distributions obtained from thermal cycle analyses at the end of the 10th cycle for $200,\mathrm{Hz}$ ((a) and (b)) and $20,\mathrm{Hz}$ ((c) and (d)). For each condition, the left panel shows the full view of the specimen with the cropped region indicated by the cyan dashed rectangle, and the right panel shows the magnified view of the cropped region.}
  \label{eqps_heatmap_10cycle}
\end{figure*}

Figure~\ref{vm_heatmap_10cycle} shows the distributions of von Mises stress at the end of the 10th cycle for (a) $200\,\mathrm{Hz}$, (b) $20\,\mathrm{Hz}$, and (c) $2\,\mathrm{Hz}$ cases.
As previously discussed by \cite{Nascimento_etal_2025}, under thermal cycle conditions with small plastic strain, the spatial heterogeneity of stress is primarily attributed to crystal orientation.
The present simulation captures variations in residual stress across different locations within the polycrystal, which qualitatively resembles the location-dependent residual stress observed in experiments \citep{Simson_2017,Santa-aho_2021}. 
However, due to the higher spatial resolution achievable in the simulations compared to typical experimental measurements, the range of stress variation may be broader than observed experimentally, as also discussed in \cite{Grilli_etal_2022}.

At $20\,\mathrm{Hz}$ and $2\,\mathrm{Hz}$, high stresses also occurred in grains near the center of the analysis domain, whereas at $200\,\mathrm{Hz}$, von Mises stress was more broadly distributed throughout the interior, excluding near the boundaries.

Figure~\ref{eqps_heatmap_10cycle} (a) and (c) show the distributions of equivalent plastic strain at the end of the 10th cycle for $200\,\mathrm{Hz}$ and $20\,\mathrm{Hz}$ cases. Figure~\ref{eqps_heatmap_10cycle} (b) and (d) present enlarged views of the cropped region indicated by the dashed rectangle in Fig.~\ref{eqps_heatmap_10cycle} (a) and (c). At $2\,\mathrm{Hz}$, the accumulated plastic strain was negligible, so the distribution is omitted here.

In the $200\,\mathrm{Hz}$ and $20\,\mathrm{Hz}$ cases, the maximum plastic strain accumulates near the boundary.
This trend is consistent with the spatial distribution of temperature amplitude, where larger thermal oscillations are imposed near the boundary region.
As a result, both the residual stress and the accumulated plastic strain become higher in these regions compared to the interior.

In thermal fatigue experiments on 316 stainless steel, crack initiation has been reported to occur preferentially at locations where the temperature amplitude is highest, such as the surface region of pipes \citep{Paffumi_2015}. 
The present simulations reproduce this tendency, showing that regions subjected to larger thermal oscillations exhibit higher stress and enhanced plastic strain accumulation, suggesting that such regions may act as potential sites of damage initiation.

Furthermore, this tendency is not merely a snapshot-specific observation as discussed in the following sections.
The spatial distribution of dominant DMD modes indicates that regions experiencing larger temperature amplitudes consistently exhibit higher stress responses over time.
This demonstrates that the observed heterogeneity is governed by persistent spatiotemporal structures rather than transient fluctuations.

There is a clear difference in the plastic strain accumulation in the cropped region.
At $200\,\mathrm{Hz}$, the equivalent plastic strain accumulated more extensively and uniformly across the wide area, whereas at $20\,\mathrm{Hz}$, the region of accumulation was more localized near the center.
Furthermore, in contrast to the fact that the maximum value of equivalent plastic strain in the entire area was higher in the $200\,\mathrm{Hz}$ case than in the $20\,\mathrm{Hz}$ case, 
the maximum value of equivalent plastic strain in the cropped region was higher at $20\,\mathrm{Hz}$ than at $200\,\mathrm{Hz}$.
The differences of equivalent plastic strain in the cropped region indicate that the $20\,\mathrm{Hz}$ case exhibits more localized plastic deformation inside the material.

For the $20\,\mathrm{Hz}$ case, relatively high von Mises stress and localized accumulation of plastic strain were observed near the center of the analysis domain at the end of the 10th cycle.
This behavior is attributed to the fact that, for each cycle, the thermal diffusion within the material responds sufficiently fast to the temperature variation imposed at the boundaries, resulting in large temperature oscillation amplitudes over a relatively wide region inside the domain.
As a consequence, the central region is constrained by the surrounding areas that undergo thermal expansion during heating, leading to stress concentration and the satisfaction of the local yield condition, which in turn causes localized plastic deformation.
In contrast, for the $200\,\mathrm{Hz}$ case, the time scale of the imposed temperature variation is much shorter than the thermal diffusion time compared to the $20\,\mathrm{Hz}$ case.
As a result, the instantaneous temperature oscillation during each cycle remains primarily localized near the boundaries.
Accordingly, regions experiencing significant thermal expansion and contraction are confined to the vicinity of the boundaries, and the resulting strain and stress are distributed more uniformly throughout the domain.
Consequently, pronounced stress concentration in the central region is suppressed, and plastic strain is distributed more uniformly, leading to response behavior that differs from that observed under the $20\,\mathrm{Hz}$ condition.

\subsection{Spatiotemporal structure of thermal stress evolution}
Before discussing the DMD results, we briefly comment on the computational efficiency of the present DMD method.

The computational time was measured using the $200\,\mathrm{Hz}$ thermal cycling condition. The thermal–crystal plasticity simulation required $59,870~\si{s}$, whereas the DMD computation required only $234~\si{s}$. This result indicates that the computational cost of DMD is significantly lower than that of the underlying thermal-mechanical simulation.

It should be noted that these results were obtained without particular optimization for computational speed. The thermal–crystal plasticity analysis was conducted using a user application built on the MOOSE framework, and the DMD was implemented using an in-house MATLAB code. Despite this, the DMD computation remains sufficiently inexpensive compared to the full-field simulation.

A breakdown of the DMD computation time shows that approximately $13~\si{s}$ was spent on singular value decomposition, and $46~\si{s}$ on the evaluation of Eqs.~\eqref{similarity}, \eqref{a9}, \eqref{exactvector}, and \eqref{alphavec}, with the remainder attributed to data input/output and matrix construction.

In this study, a greedy algorithm is employed to reorder the DMD modes based on reconstruction error. This procedure is an additional post-processing step for mode interpretation and is not required for standard DMD reconstruction; therefore, its computational cost is not included in the above estimate.

\subsubsection{DMD reconstruction error analysis}

\begin{figure*}[t]   
  \centering
  \includegraphics[width=\textwidth]{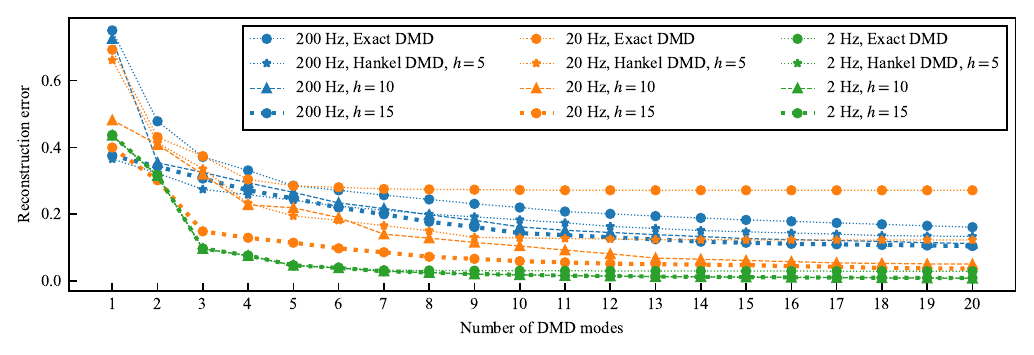}
  \caption{Reconstruction error versus the number of DMD modes for three thermal cycle frequencies: $200\,\mathrm{Hz}$, $20\,\mathrm{Hz}$, and $2\,\mathrm{Hz}$. The results are shown for greedy-ordered DMD modes.}
  \label{reconstruction_error}
\end{figure*}

Figure~\ref{reconstruction_error} shows the reconstruction error defined in Eq.~\eqref{recon_error_def} when applying Exact DMD and Hankel DMD with time-delay embedding dimensions of $h=5, 10,$ and $15$ to the thermal-crystal plasticity simulation results at $200\,\mathrm{Hz}$, $20\,\mathrm{Hz}$, and $2\,\mathrm{Hz}$. 
The horizontal axis represents the number of modes used for reconstruction, with DMD modes sorted by importance using a greedy algorithm.
Consistent with the previous report \citep{Asada_Kawai_2024}, increasing the time-delay dimension $h$ improved the reconstruction error.

Here, the sensitivity of the reconstruction error to the time-delay embedding dimension $h$ is interpreted as reflecting the strength of history dependence and nonlinear effects in the system.
Particularly for the $20\,\mathrm{Hz}$ case, while the reconstruction error using 20 modes with Exact DMD was approximately $\epsilon^{(1,20)}=0.3$, Hankel DMD showed significant improvement: about $\epsilon^{(5,20)}=0.15$ for $h=5$, $\epsilon^{(10,20)}=0.09$ for $h=10$, and $\epsilon^{(15,20)}=0.08$ for $h=15$. The improvement in the reconstruction error in increasing $h$ can be interpreted as the Hankel DMD with $h$ larger than one, effectively capturing the system's history dependence by introducing a time-delay embedding dimension.
On the other hand, in the $2\,\mathrm{Hz}$ case, where plastic strain accumulation was negligible, the system exhibited minimal history dependence, as expected since the time-series waveform showed a monotonically periodic response with nearly identical amplitudes.
Therefore, the reduction in error from increasing the time-delay embedding dimension $h$ was not as pronounced as in the other two cases.

Using $h=15$ suppressed the reconstruction error to approximately $0.1$ in all cases.
While using a larger $h$ might further reduce the error, the computational cost and size of the data matrix increase sharply with $h$ in Hankel DMD.
Therefore, $h=15$ is used as the representative value for subsequent discussions on the DMD mode structure and spectral properties.

\subsubsection{DMD modal analysis}

\begin{figure*}[htbp]   
  \centering
  \includegraphics[width=\textwidth]{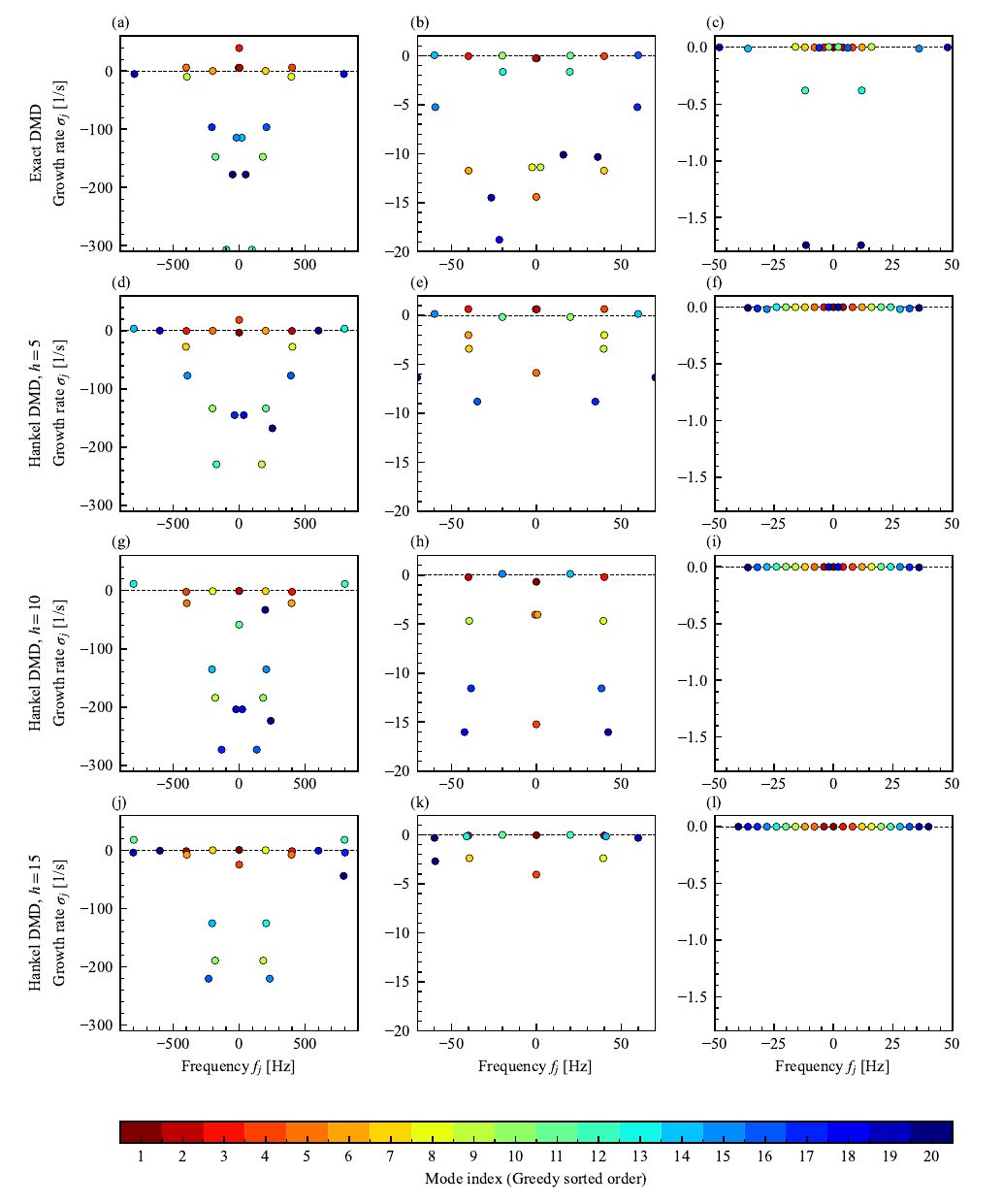}
  \caption{Continuous-time eigenvalue spectra obtained by DMD. Figure (a), (d), (g), and (j) correspond to DMD results for thermal cycle frequency of $200\,\mathrm{Hz}$. Figure (b), (e), (h), and (k) correspond to the thermal cycle frequency of $20\,\mathrm{Hz}$. Figure (c), (f), (i), and (l) correspond to the thermal cycle frequency of $2\,\mathrm{Hz}$. Each row corresponds to the different Hankel dimensions $h$. The frequency $f_{j}$ is shown on the horizontal axis and the growth rate $\sigma_{j}$ on the vertical axis. The color encodes the greedy-sorted mode index (top 20 modes). A dashed line at zero growth rate is shown for reference.}
  \label{dmd_freq_growth_plot}
\end{figure*}

Figure~\ref{dmd_freq_growth_plot} shows the continuous eigenvalue spectra obtained by DMD for three thermal cycle frequencies of $200\,\mathrm{Hz}$, $20\,\mathrm{Hz}$, and $2\,\mathrm{Hz}$.
The vertical axis represents the growth rate $\sigma_{j}$ defined in Eq.~\eqref{freqdef}, corresponding to the real part of the continuous eigenvalue.
The horizontal axis represents the modal frequency $f_{j}$ calculated from the imaginary part of the continuous eigenvalue based on Eq.~\eqref{freqdef}. 
Negative frequencies arise from complex-conjugate mode pairs and correspond to the same physical oscillatory modes; the magnitude $|f_{j}|$ characterizes the oscillation frequency. 
Each marker is colored by the greedy importance order, and only the first 20 modes are shown.

As shown in Fig.~\ref{dmd_freq_growth_plot}(a), (d), (g), and (j) and Fig.~\ref{dmd_freq_growth_plot}(b), (e), (h), and (k), in the case of $200\,\mathrm{Hz}$ and $20\,\mathrm{Hz}$, the position of plots is changed with the changes of $h$.
As the time-delay embedding dimension $h$ increases, additional modes with distinct frequencies and growth rates emerge that are not captured for smaller values of $h$.
The emergence of mode in increasing $h$ indicates that incorporating time-delay dimensions allows the extraction of history-dependent dynamics that cannot be represented by instantaneous state information alone.
On the other hand, the distribution of modes with growth rates close to zero remains unchanged mainly with respect to $h$.
These near-zero-growth modes therefore represent quasi-steady or persistent spatiotemporal response structures that are robust to the choice of the time-delay dimension and are likely to characterize the dominant behavior of the system under the present conditions.

In contrast, as shown in Fig.~\ref{dmd_freq_growth_plot}(c), (f), (i), and (l), the $2\,\mathrm{Hz}$ case showed almost the same plot position across the different $h$ as the former two cases, which is consistent with a steady-state periodic response with weak history dependence.
The following discussion is based on the DMD result for $h=15$.

\begin{table}[t]
\centering
\caption{DMD summary for $200\,\mathrm{Hz}$ case, Hankel DMD ($h=15$)}
\begin{tabular}{rrr}
\hline
Mode & $f_{j}$ [Hz] & $\sigma_{j}$ [1/s] \\
\hline
  1 & $    0.0000$ & $    0.606995$ \\
  2 & $ -400.6745$ & $   -1.557586$ \\
  3 & $  400.6745$ & $   -1.557586$ \\
  4 & $    0.0000$ & $  -24.730774$ \\
  5 & $  396.3692$ & $   -7.689255$ \\
  6 & $ -396.3692$ & $   -7.689255$ \\
  7 & $ -200.2323$ & $   -0.034973$ \\
  8 & $  200.2323$ & $   -0.034973$ \\
  9 & $  182.0678$ & $ -189.444933$ \\
 10 & $ -182.0678$ & $ -189.444933$ \\
 11 & $ -797.9324$ & $   17.925467$ \\
 12 & $  797.9324$ & $   17.925467$ \\
 13 & $  204.3650$ & $ -125.409000$ \\
 14 & $ -204.3650$ & $ -125.409000$ \\
 15 & $  231.4319$ & $ -220.552834$ \\
 16 & $ -231.4319$ & $ -220.552834$ \\
 17 & $  801.5548$ & $   -4.030958$ \\
 18 & $  600.1081$ & $   -0.711429$ \\
 19 & $ -801.5548$ & $   -4.030958$ \\
 20 & $ -600.1081$ & $   -0.711429$ \\
\hline
\end{tabular}
\label{table_200}
\end{table}

\begin{table}[t]
\centering
\caption{DMD summary for $20\,\mathrm{Hz}$ case, Hankel DMD ($h=15$)}
\begin{tabular}{rrr}
\hline
Mode & $f_{j}$ [Hz] & $\sigma_{j}$ [1/s] \\
\hline
  1 & $    0.0000$ & $   -0.026498$ \\
  2 & $   39.9991$ & $   -0.069252$ \\
  3 & $  -39.9991$ & $   -0.069252$ \\
  4 & $    0.0000$ & $   -4.052703$ \\
  5 & $   79.9992$ & $   -0.395570$ \\
  6 & $  -79.9992$ & $   -0.395570$ \\
  7 & $  -39.4102$ & $   -2.378897$ \\
  8 & $   39.4102$ & $   -2.378897$ \\
  9 & $  -79.3792$ & $   -2.456624$ \\
 10 & $   79.3792$ & $   -2.456624$ \\
 11 & $  -19.9883$ & $    0.000823$ \\
 12 & $   19.9883$ & $    0.000823$ \\
 13 & $  -41.0718$ & $   -0.152438$ \\
 14 & $   41.0718$ & $   -0.152438$ \\
 15 & $  120.0259$ & $   -0.651542$ \\
 16 & $  119.3980$ & $   -1.630653$ \\
 17 & $ -120.0259$ & $   -0.651542$ \\
 18 & $ -119.3980$ & $   -1.630653$ \\
 19 & $   59.9711$ & $   -0.304171$ \\
 20 & $  -59.9711$ & $   -0.304171$ \\
\hline
\end{tabular}
\label{table_20}
\end{table}

\begin{table}[t]
\centering
\caption{DMD summary for $2\,\mathrm{Hz}$ case, Hankel DMD ($h=15$)}
\begin{tabular}{rrr}
\hline
Mode & $f_{j}$ [Hz] & $\sigma_{j}$ [1/s] \\
\hline
  1 & $    0.0000$ & $    0.000000$ \\
  2 & $   -4.0000$ & $    0.000005$ \\
  3 & $    4.0000$ & $    0.000005$ \\
  4 & $    8.0000$ & $    0.000009$ \\
  5 & $   -8.0000$ & $    0.000009$ \\
  6 & $  -12.0000$ & $    0.000006$ \\
  7 & $   12.0000$ & $    0.000006$ \\
  8 & $   16.0000$ & $   -0.000016$ \\
  9 & $  -16.0000$ & $   -0.000016$ \\
 10 & $   20.0000$ & $   -0.000085$ \\
 11 & $  -20.0000$ & $   -0.000085$ \\
 12 & $   23.9999$ & $   -0.000226$ \\
 13 & $  -23.9999$ & $   -0.000226$ \\
 14 & $   27.9999$ & $   -0.000432$ \\
 15 & $  -27.9999$ & $   -0.000432$ \\
 16 & $   31.9999$ & $   -0.000651$ \\
 17 & $  -31.9999$ & $   -0.000651$ \\
 18 & $  -35.9998$ & $   -0.000746$ \\
 19 & $   35.9998$ & $   -0.000746$ \\
 20 & $  -39.9998$ & $   -0.000246$ \\
\hline
\end{tabular}
\label{table_2}
\end{table}

Tables~\ref{table_200},~\ref{table_20},~and~\ref{table_2} summarize the leading DMD modes for thermal cycle periods of $200\,\mathrm{Hz}$, $20\,\mathrm{Hz}$, and $2\,\mathrm{Hz}$. 
The table lists the modal frequency $f_{j}$ and the growth (decay) rate $\sigma_{j}$, corresponding to the plot shown in Fig.~\ref{dmd_freq_growth_plot}. 
Only the first 20 modes are reported.
For $200\,\mathrm{Hz}$ and $20\,\mathrm{Hz}$, most of the extracted modes have negative growth rates, meaning their responses gradually decay, and suggest convergence toward a periodic response.
In contrast, the case of $2\,\mathrm{Hz}$ is mainly governed by the mode with a growth rate of about zero.
The presence of a dominant mode with an approximately zero growth rate indicates that the stress response is primarily governed by a stable mode, corresponding to periodic steady-state behavior without noticeable transient decay or growth.
This DMD-based interpretation is consistent with the nearly perfectly periodic stress response identified in the time-series analysis.
The amplitude of the growth rate is largest in the order of $200\,\mathrm{Hz}$, $20\,\mathrm{Hz}$, and $2\,\mathrm{Hz}$, confirming that higher frequencies exhibit stronger non-stationary behavior.

Focusing on the frequencies shown in Tables~\ref{table_200},~\ref{table_20},~and~\ref{table_2}, the degree of synchronization with the external loading frequency was evaluated using the relative deviation from the nearest integer multiple of the thermal cycle frequency.
The relative deviation is defined as
\begin{equation}
    \frac{|f_{j}-n\cdot f|}{f},
    \label{deviation}
\end{equation}
where $f_{j}$ is the modal frequency calculated by Eq.~\eqref{freqdef}, $f$ is the thermal cycle frequency, and $n$ is selected to minimize the value of Eq.~\eqref{deviation}.
For the $2\,\mathrm{Hz}$ case, all modes exhibit very small relative deviations, with the maximum value remaining below 0.01, indicating that the system response is strongly locked to the imposed periodic loading.
In the $20\,\mathrm{Hz}$ case, although most modes still cluster around integer multiples of the thermal cycle frequency, the maximum relative deviation increases to approximately 0.027, suggesting a moderate relaxation of the strict harmonic constraint.
The $200\,\mathrm{Hz}$ also has some modes whose frequency shows a breakdown of the integer-multiple structure. The maximum relative deviation is 0.16, clearly indicating the emergence of inharmonic modes that are not directly synchronized with the thermal cycle frequency.

The progressive increase in the relative frequency deviation with increasing thermal cycle frequency indicates that, under high-frequency thermal cycle conditions, the system response cannot be represented solely by harmonic components synchronized with the external loading. 
This behavior reflects the increased complexity and nonlinearity of the thermal-mechanical response captured in the CPFEM simulations.

It should be emphasized that the DMD modes provide a data-driven decomposition of the simulated dynamics based on a linear representation in an augmented state space. 
Therefore, the extracted modal frequencies should be interpreted as effective descriptors of the observed system behavior, rather than direct measures of intrinsic material time scales. 
They instead reflect dominant temporal structures emerging from the coupled thermal-mechanical response captured in the simulations.

From the above analyses, the low-frequency cycle ($2\,\mathrm{Hz}$) exhibits behavior close to a steady-state periodic response, intermediate frequency cycle ($20\,\mathrm{Hz}$) begins to show history dependence, and high-frequency cycle ($200\,\mathrm{Hz}$) is characterized by strong nonlinearity and inharmonicity.
These trends are consistent with the above discussion based on the time-delay embedding dimension $h$-dependence of the reconstruction error shown in Fig.~\ref{reconstruction_error}, providing eigenvalue-based evidence that the effectiveness of time-delay embedding increases with thermal cycle frequency.

Overall, the DMD results demonstrate that the complex spatiotemporal behavior obtained from thermal-crystal plasticity simulations can be systematically decomposed into a small number of dominant modes. 
This capability provides a structured way to interpret high-dimensional simulation data and to distinguish persistent response patterns from transient features.

In this sense, DMD serves as an effective post-processing and diagnostic tool for analyzing thermal-mechanical responses under cyclic loading, rather than a means to directly identify new physical mechanisms.
Accordingly, the extracted DMD modes should be interpreted as effective descriptors of the spatiotemporal structure of the simulated stress fields, rather than as intrinsic.

\subsubsection{Spatial structure of DMD mode}

\begin{figure*}[htbp]   
  \centering
  \includegraphics[width=\textwidth]{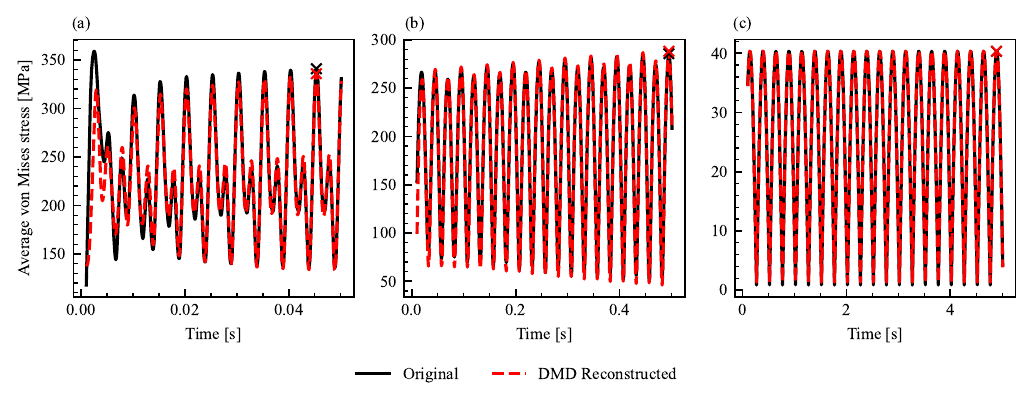}
  \caption{Comparison of the temporal evolution of the spatially averaged von Mises stress between the original snapshots and the DMD-reconstructed data for three datasets corresponding to thermal cycle frequencies of $200\,\mathrm{Hz}$, $20\,\mathrm{Hz}$, and $2\,\mathrm{Hz}$. DMD reconstruction employs the top 20 modes with Hankel dimension $h=15$. Cross markers indicate the time steps corresponding to the maximum stress in the final cycle, which are used in Fig.~\ref{dmd_vector_visualization} to visualize the distribution of DMD modes. Panels (a), (b), and (c) correspond to datasets with increasing thermal cycle frequency.}
  \label{recon_mean_vm}
\end{figure*}

Figure~\ref{recon_mean_vm} shows the temporal evolution of the spatially averaged von Mises stress for each thermal cycle frequency ($200\,\mathrm{Hz}$, $20\,\mathrm{Hz}$, and $2\,\mathrm{Hz}$). 
The solid black lines represent the original thermal-crystal plasticity simulation results, while the dotted red lines denote the Hankel DMD reconstructions obtained using $h=15$ and the top 20 modes.
Overall, the DMD reconstruction captures the main features of the original waveforms, although a slight underestimation of peak stress is observed during the initial transient cycles at $200\,\mathrm{Hz}$ and $20\,\mathrm{Hz}$.
In contrast, the reconstruction accuracy at $2\,\mathrm{Hz}$ is high due to the nearly identical periodic response.

\begin{sidewaysfigure*}   
  \centering
  \includegraphics[width=\textwidth]{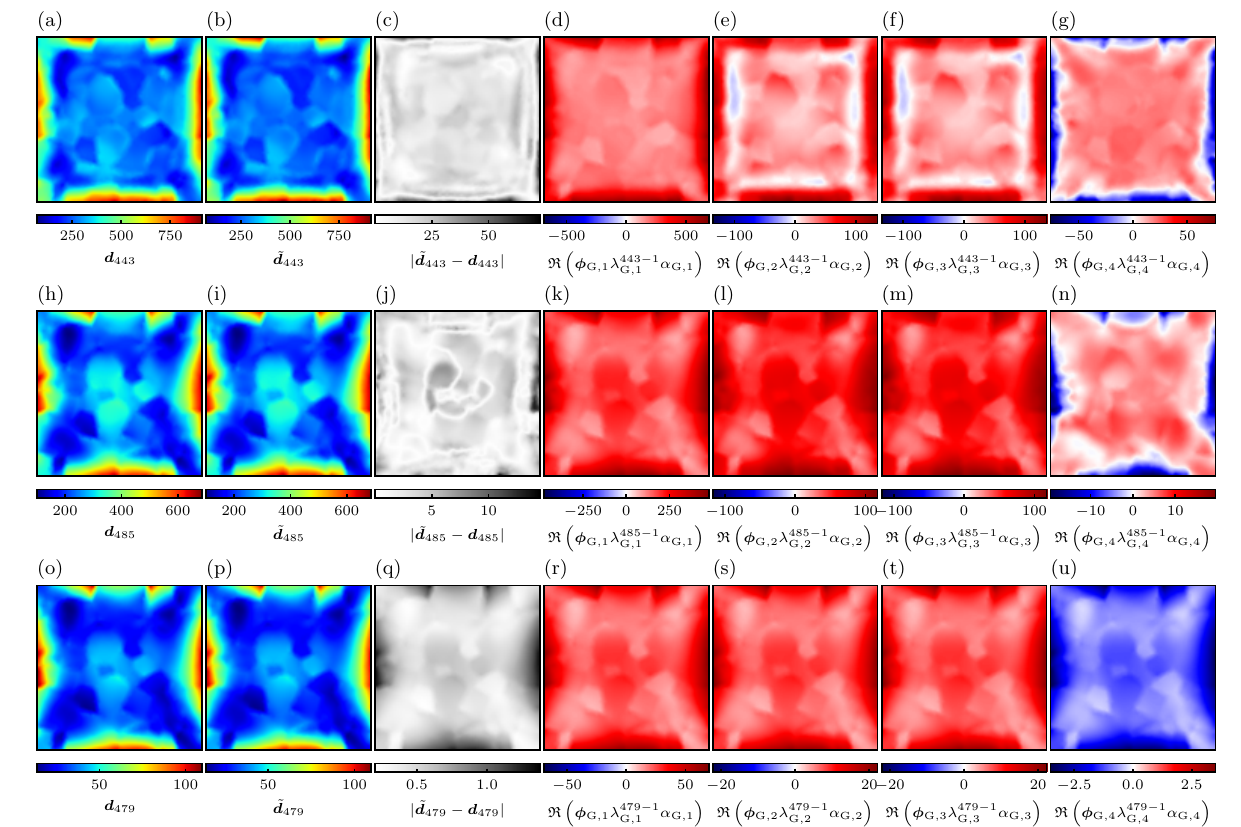}
  \caption{Spatial distribution of von Mises stress reconstructed by DMD for three datasets corresponding to thermal cycle frequencies of $200\,\mathrm{Hz}$, $20\,\mathrm{Hz}$, and $2\,\mathrm{Hz}$ (top to bottom). from left to right, each column shows: the original snapshot (Figure (a), (h), and (o)), the Hankel DMD ($h=15$) reconstruction using the top 20 modes (Figure (b), (i), and (p)) , the absolute reconstruction error (Figure (c), (j), and (q)), and the contributions of first four modes of greedy sorted DMD modes (Figure (d), (e), (f), (g), (k), (l), (m), (n), (r), (s), (t), and (u)). The snapshots correspond to the time of maximum stress in the final cycle, which is indicated by cross marker in Fig.~\ref{recon_mean_vm}.}
  \label{dmd_vector_visualization}
\end{sidewaysfigure*}

Figure~\ref{dmd_vector_visualization} shows the von Mises stress distribution at the time indicated by the cross mark in Fig.~\ref{recon_mean_vm} for each thermal cycle frequency ($200\,\mathrm{Hz}$, $20\,\mathrm{Hz}$, and $2\,\mathrm{Hz}$), the von Mises stress distribution reconstructed using the top 20 DMD modes, the error distribution, and the distributions of the real part of the top four modes.
Figure~\ref{dmd_vector_visualization}(e) and (f), (l) and (m), and (s) and (t) showed the same distribution because these mode pairs correspond to complex-conjugate eigenvalues, as listed in Tables~\ref{table_200}, \ref{table_20}, and \ref{table_2}, and therefore share the same spatial mode shape.
The three left columns of images indicate that the reconstruction using 20 DMD modes reproduces the von Mises stress distributions of the original thermal-crystal plasticity simulation results well, indicating that DMD can extract the governing spatiotemporal structure with a small amount of data.

In all cases, the first-order mode is the stationary mode whose frequency is $0\,\mathrm{Hz}$.
At the $200\,\mathrm{Hz}$ case shown in Fig.~\ref{dmd_vector_visualization}(d), the growth rate is a positive value of $0.606995$, implying that the presence of non-stationary behavior increases the stress level according to the distribution of the shown eigenvector.
The growth rate for $20\,\mathrm{Hz}$ shown in Fig.~\ref{dmd_vector_visualization}(k) and $2\,\mathrm{Hz}$ shown in Fig.~\ref{dmd_vector_visualization}(r) cases are about $-0.026498$ and $0.000000$, indicating that the steady-state dominant structure prevails at lower frequencies.
The first-order modes at $20\,\mathrm{Hz}$ and $2\,\mathrm{Hz}$ show high contributions from grains near the center, reflecting long-term trends in stress concentration.
This DMD-based observation allows us to avoid misinterpreting the distributions in Fig.~\ref{vm_heatmap_10cycle}(b) and (c) as "random local distribution" in the single snapshot observation performed earlier, demonstrating that DMD is useful for capturing dominant structures.

The second mode corresponds to approximately twice the frequency of the applied thermal loading in each case.
At $200\,\mathrm{Hz}$ it is $400.6745\,\mathrm{Hz}$, at $20\,\mathrm{Hz}$ it is $39.9991\,\mathrm{Hz}$ and at $2\,\mathrm{Hz}$ it is $4.0000\,\mathrm{Hz}$, with growth rates of $-1.557586$, $-0.069252$, and $0.000005$, respectively.
At $200\,\mathrm{Hz}$ shown in Fig.~\ref{dmd_vector_visualization}(e) and (f), the spatial distributions of mode $\bm{\phi}_{\mathrm{G},2}\lambda^{443-1}_{\mathrm{G},2}\alpha_{\mathrm{G},2}$ and $\bm{\phi}_{\mathrm{G},3}\lambda^{443-1}_{\mathrm{G},3}\alpha_{\mathrm{G},3}$ exhibits both positive and negative regions.
The exhibition of both positive and negative regions indicates a spatial phase variation in the stress response, suggesting that the thermal-mechanical response is no longer homogeneous oscillation and that local regions respond with different phases to the thermal loading.
Such behavior reflects the emergence of inharmonic and non-synchronous stress oscillations during a high-frequency thermal cycle.
In contrast, for the $20\,\mathrm{Hz}$ shown in Fig.~\ref{dmd_vector_visualization}(l) and (m) and $2\,\mathrm{Hz}$ shown in Fig.~\ref{dmd_vector_visualization}(s) and (t), the mode distributions are dominated by a single sign over most of the domain, implying a nearly in-phase oscillatory response across the material.

The third mode has a frequency of $0.0000\,\mathrm{Hz}$ at $200\,\mathrm{Hz}$ and $20\,\mathrm{Hz}$ cases, indicating it is associated with a slowly varying background stress field.
Unlike the first mode, these modes exhibit the sign reversals and possess significantly larger decay rates ($-24.730774$ and $-4.052703$).
The strongly decaying behavior suggests that the third mode primarily represents a transient adjustment of the spatial stress distribution during the early cycles, rather than a persistent long-term trend.
In combination with the first stationary mode, it contributes to shaping the evolving stress heterogeneity during the transient regime, after which its influence rapidly diminishes.
In the $2\,\mathrm{Hz}$ case, the first four modes shown in Fig.~\ref{dmd_vector_visualization}(r), (s), (t), and (u) exhibit very similar spatial distributions, indicating that a simple, steady-state structure dominates the system response.
Although sign differences are observed among the modes, their spatial patterns are nearly identical, indicating the absence of pronounced transient or inharmonic behavior under a low-frequency thermal cycle.

The results above confirm that DMD is effective for identifying a low-dimensional representation of the spatiotemporal structures present in the simulation data and for quantitatively understanding their relationship with frequency, growth rate, and distribution of eigenvectors.
Specifically, inharmonic structures strongly emerge under high-frequency thermal cycles, whereas steady-state structures prevail under low-frequency thermal cycles.
The above investigations demonstrate that DMD mode contribution analysis enables clear evaluation of long-term trends and dominant structures that are often overlooked in single-snapshot observations.

\subsection{Long-term prediction performance and stability}

\begin{figure*}[htbp]   
  \centering
  \includegraphics[width=\textwidth]{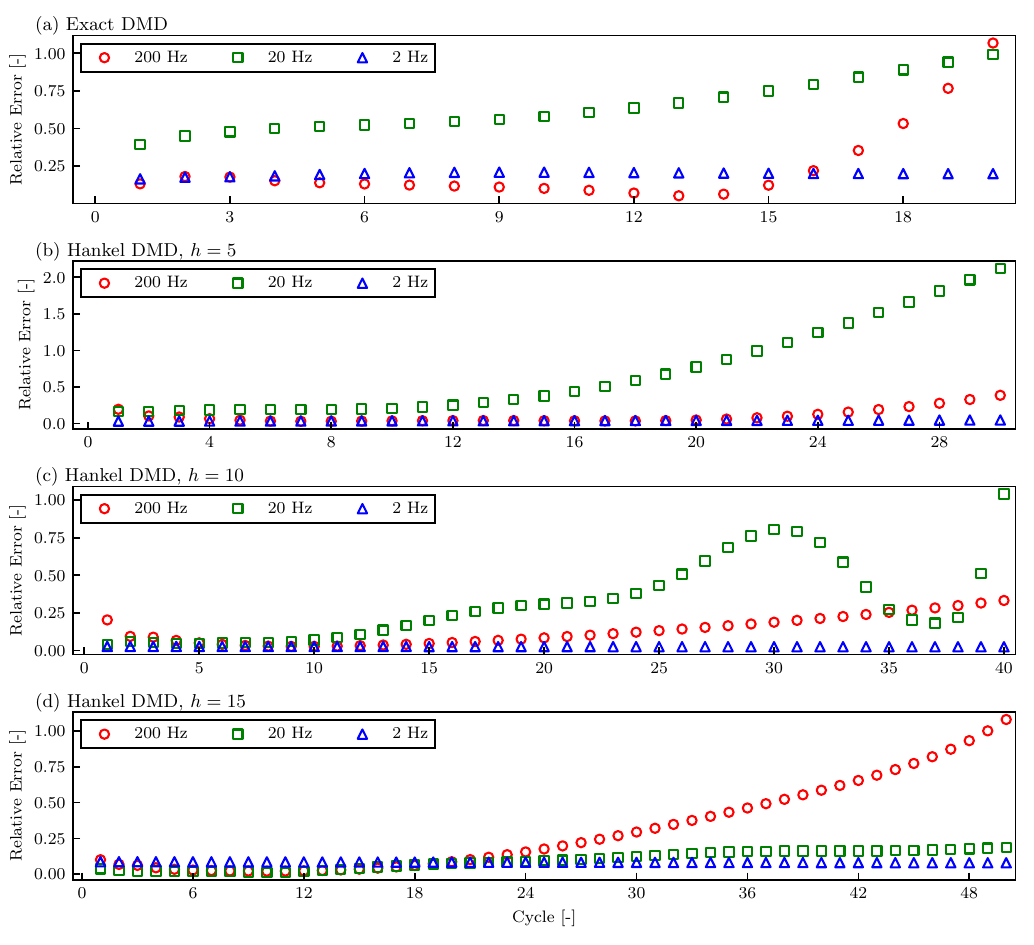}
  \caption{Cycle-wise evolution of the relative reconstruction error obtained by DMD for three thermal cycle frequencies ($200\,\mathrm{Hz}$, $20\,\mathrm{Hz}$, and $2\,\mathrm{Hz}$). Each panel corresponds to a Hankel embedding dimension $h=1,5,10$, and $15$ from top to bottom. Markers denote different thermal cycle frequencies, and the error is evaluated using the top 20 DMD modes. White markers are used to improve visibility where data points overlap.}
  \label{dmd_prediction}
\end{figure*}

Figure~\ref{dmd_prediction} shows the cycle-wise evolution of the relative reconstruction error of the von Mises stress field.
Each reconstruction is based on the top 20 DMD modes obtained with different Hankel embedding dimensions $h$.
Cycles beyond the 11th correspond to extrapolated predictions.
The relative error is defined as
\begin{equation}
    \epsilon_{\mathrm{cycle},i}=\frac{\lvert \tilde{\bm{d}}_{i} -\bm{d}_{i}\rvert}{\lvert\bm{d}_{i}\rvert}.
\end{equation}

Since the DMD modes are selected using a greedy algorithm that minimizes reconstruction error over the entire Hankel matrix, direct comparisons of absolute prediction error among different values of $h$ should be treated with caution.
Nevertheless, the temporal evolution of the prediction error within each condition, as well as the relative trends among different thermal cycle frequencies, provides meaningful insight into the long-term prediction capability of DMD.

All values of time-delay embedding dimension $h$ exhibit consistently stable long-term prediction accuracy for $2\,\mathrm{Hz}$, compared to the other thermal cycle frequencies.
This likely reflects the difference in time-dependent behavior between thermoelastic and thermoelastoplastic responses. 
When the elastic response dominates, the time evolution of the stress field is relatively linear and reversible, and tends to be well predicted using a few DMD modes. 
On the other hand, under thermoelastic-plastic conditions, the time evolution of the stress field exhibits stronger history dependence due to the accumulation of plastic deformation and the history dependence of internal state variables. 
This nonlinear and history-dependent behavior complicates prediction with a limited set of DMD modes, resulting in increased prediction errors.

For $2\,\mathrm{Hz}$, 
the prediction errors at the end of the 15th cycle for Exact DMD and Hankel DMD with $h=5$, $10$, and $15$ were $0.20$, $0.037$, $0.027$, and $0.064$, respectively.
The increase in error from $h=10$ to $h=15$ suggests that an excessively large Hankel dimension may become counterproductive for such a simple system, as the additional degrees of freedom do not improve predictive performance.

In Fig.~\ref{dmd_prediction}(c), the relative error for $20\,\mathrm{Hz}$ exhibits a local maximum around the 30th cycle, a behavior not observed at $2\,\mathrm{Hz}$ or $200\,\mathrm{Hz}$. 
Oscillatory behavior in DMD reconstruction error is generally expected due to mode truncation, phase shifts, and mode interference. 
Since the error is evaluated only at the end of each cycle, high-frequency oscillations within a cycle are not visible. 
Therefore, the local peak at $20\,\mathrm{Hz}$ represents a low-frequency effect, while similar high-frequency variations may exist for the other frequencies but are not resolved. 
Limiting the number of modes for reconstruction may have amplified these effects, causing a temporary increase in error. 
In contrast, at $200,\mathrm{Hz}$, the presence of DMD modes with large growth rates dominates the error increase due to their exponential amplification, leading to a more monotonically increasing error rather than a locally oscillatory one.

At $20\,\mathrm{Hz}$, increasing the Hankel dimension clearly improves the prediction accuracy and stabilizes long-term predictions.
Comparing cycles where prediction error first exceeds $0.2$ after the training window, the corresponding cycles are the $11$th cycle (with Exact DMD, where the error already exceeds $0.2$ from the first cycle), the $11$th cycle ($h=5$), the $15$th cycle ($h=10$), and the $53$rd cycle ($h=15$). 
This trend indicates that incorporating a larger time-delay embedding dimension effectively captures the history dependence of the system, thereby enhancing long-term prediction performance.

At $200\,\mathrm{Hz}$, the prediction error first exceeds $0.2$ at 16th, 27th, 32nd, and 27th cycles for Exact DMD and Hankel DMD with $h=5$, $10$, and $15$, respectively.
Once the prediction error for $h=15$ exceeds that for $h=10$, the discrepancy continues to increase as extrapolation proceeds.
This result demonstrates that increasing the Hankel dimension improves short-term reconstruction accuracy but does not necessarily lead to monotonically improved long-term prediction stability. 
In the high-frequency case of $200\,\mathrm{Hz}$, previous analyses suggest that strong material nonlinearity and complex transient behavior significantly influence the system evolution.
As a result, linear approximation through DMD becomes increasingly sensitive to the choice of embedding dimension and mode selection.
Furthermore, in the continuous eigenvalue spectrum shown in Fig.~\ref{dmd_freq_growth_plot}, the number of modes with positive growth rates at $h=15$ increased to three, compared to two at $h=10$, which may contribute to error amplification during long-term prediction.

Overall, these results indicate that the long-term forecast performance of DMD depends strongly on both the frequency characteristics of the target system and the prediction timescale.
The choice of Hankel embedding dimension involves a trade-off between short-term reconstruction accuracy and long-term prediction stability.
This highlights the importance of carefully selecting model parameters when applying DMD to nonlinear thermal-mechanical systems, particularly under high-frequency loading conditions.

\section{Conclusion}
\label{Conclusion}
In this study, thermal-crystal plasticity simulations were conducted under different frequency conditions to investigate their influence on the spatiotemporal structure of thermal stress fields and plastic responses within polycrystalline microstructure. 
The main conclusions of this study are summarized as follows:

\begin{enumerate}
\item
The influence of thermal cycle frequency on the internal temperature distribution and resulting stress fields was examined. 
As expected from classical heat conduction scaling, the transition between quasi-steady and unsteady thermal–mechanical behavior can be predicted using the Fourier number: 
for large Fourier numbers (e.g., $\mathrm{Fo}\approx 2.5$), the temperature field approaches a quasi-steady state, whereas for small Fourier numbers (e.g., $\mathrm{Fo}\approx 0.025$ and $0.25$), strong spatial temperature gradients persist. 
This confirms that the interaction between thermal cycle frequency and microstructural length scales governs the overall thermal–plastic response.

\item
By applying DMD as a diagnostic technique for thermal-crystal plasticity simulation, it was demonstrated that the temporal evolution of the thermal stress field can be represented by a limited number of spatiotemporal modes that provide a compact representation of the simulated stress field evolution within the simulation data, and that the structure of these modes strongly depends on the thermal cycle frequency.

\item
Using the extracted DMD modes, predictions of residual stress were performed in regions beyond the cycle range used to construct the modes. 
The results indicate that, while DMD provides stable and accurate predictions for predominantly thermoelastic responses, the prediction accuracy for thermoelastoplastic responses becomes increasingly unstable with the number of cycles. 
Moreover, the time-delay embedding dimension $h$ governs a trade-off between short-term reconstruction accuracy and long-term prediction stability, highlighting the need to carefully select this hyperparameter when applying DMD to nonlinear thermal–mechanical systems.
It should be emphasized that the DMD modes extracted in this study represent a low-dimensional representation of the simulated dynamics, rather than uniquely identifiable physical mechanisms inherent to the material itself.

\item
Overall, this study demonstrates that combining CPFEM with DMD provides a useful framework for analyzing and organizing complex thermal-mechanical simulation data under cyclic thermal loading, offering a complementary perspective to conventional analysis approaches.
\end{enumerate}

\section*{CRediT authorship contribution statement}
Haruki Ohashi: Conceptualization, Data curation, Formal analysis, Investigation, Methodology, Software, Validation, Visualization, Writing - original draft.

Yoshiteru Aoyagi: Project administration, Resources, Supervision, Writing - review \& editing.

\section*{Declaration of competing interests}
The authors declare the following financial interests/personal relationships which may be considered as potential competing interests: Haruki Ohashi reports financial support was provided by the Advanced Graduate School Research Initiative for International Scholarly Excellence (AGS RISE) Program, Tohoku University.

\section*{Acknowledgement}
This work was supported by AGS RISE Program, Tohoku University.
Part of this work was conducted under the framework of the Graduate Program for Integration of Mechanical Systems (GP-Mech), Tohoku University.

\section*{Data availability}
No data was used for the research described in the article.

\appendix
\section{Detailed derivations of DMD}
\label{app1}
This appendix summarizes the detailed derivation of the Exact DMD formulation \citep{Tu_etal_2014,Brunton_Kutz_2019} used in Section~\ref{dmd}.

The same notation is used as in Section~\ref{dmd}, so $\bm{d}_{k}$ represents the data vector at the $k$-th time step, where $k=1,2,\dots,n$, and $n$ is the total number of time steps.
The constant matrix $\bm{A}$ is assumed to satisfy the following relation:
\begin{equation}
    \bm{d}_{k+1}\approx\bm{A}\bm{d}_{k}\quad(k=1,2,\dots,n-1).
\end{equation}
Here, the following matrices are introduced by arranging vectors $\bm{d}_{k}$:
\begin{equation}
    \bm{D}_{1}=[\bm{d}_{1},\bm{d}_{2},...,\bm{d}_{n-1}]
    ,\quad
    \bm{D}_{2}=[\bm{d}_{2},\bm{d}_{3},...,\bm{d}_{n}].
    \label{datamatrix}
\end{equation}
Since $\bm{A}$ is constant and independent of $k$, the following relation holds:
\begin{equation}
    \bm{D}_{2}\approx\bm{A}\bm{D}_{1}.
    \label{dmd_start}
\end{equation}

Using the known matrices $\bm{D}_{1}$ and $\bm{D}_{2}$, the eigenvalues and eigenvectors of $\bm{A}$ can be approximated as follows.
First, we define the objective function $\mathcal{J}$ and determine the matrix $\bm{A}$ that minimizes it:
\begin{equation}
    \mathcal{J}=\Vert \bm{D}_{2}-\bm{A}\bm{D}_{1} \Vert_{\mathrm{F}}, 
    \label{object2}
\end{equation}
where $\Vert\circ\Vert_{\mathrm{F}}$ denotes the Frobenius norm.
The matrix $\bm{A}$ that minimizes the objective function $\mathcal{J}$ can be obtained by
\begin{equation}
    \bm{A}=\bm{D}_{2}\bm{D}^{\dagger}_{1},
    \label{aequal2}
\end{equation}
where $^{\dagger}$ denotes the Moore-Penrose pseudo-inverse.

The Moore-Penrose pseudo-inverse $\bm{D}^{\dagger}_{1}$ can be obtained by using the singular value decomposition
\begin{equation}
    \bm{D}_{1}=\widetilde{\bm{U}}\widetilde{\bm{\mathit{\Sigma}}}\widetilde{\bm{W}}^{\mathsf{H}}
\end{equation}
as 
\begin{equation}
    \bm{D}^{\dagger}_{1}=\widetilde{\bm{W}}\widetilde{\bm{\mathit{\Sigma}}}^{-1}\widetilde{\bm{U}}^{\mathsf{H}}.
\end{equation}
Here, $^\mathsf{H}$ denotes the Hermitian transpose, and the matrices $\widetilde{\bm{U}}$ and $\widetilde{\bm{W}}$ satisfy $\widetilde{\bm{U}}^{\mathsf{H}}\widetilde{\bm{U}}=\bm{I}$ and $\widetilde{\bm{W}}^{\mathsf{H}}\widetilde{\bm{W}}=\bm{I}$.

Direct computation of the eigenvalues of $\bm{A}$ is computationally expensive. Therefore, we perform a similarity transformation on $\bm{A}$ using the left singular matrix $\widetilde{\bm{U}}$ obtained from the previous low-rank approximation, as follows:
\begin{equation}
    \widetilde{\bm{A}}=\widetilde{\bm{U}}^{\mathsf{H}}\bm{A}\widetilde{\bm{U}}=\widetilde{\bm{U}}^{\mathsf{H}}\bm{D}_{2}\widetilde{\bm{W}}\widetilde{\bm{\mathit{\Sigma}}}^{-1}.
    \label{similarity}
\end{equation}
Since the eigenvalues of $\widetilde{\bm{A}}$ are equal to those of $\bm{A}$, we perform eigenvalue decomposition on $\widetilde{\bm{A}}$ as
\begin{equation}
    \widetilde{\bm{A}}=\bm{V}\bm{\mathit{\Lambda}}\bm{V}^{-1},
    \label{a9}
\end{equation}
where $\bm{\mathit{\Lambda}}$ is the diagonal matrix, whose components are eigenvalues of $\widetilde{\bm{A}}$ and $\bm{V}$ is the matrix whose column vectors are eigenvectors of $\widetilde{\bm{A}}$.

\cite{Tu_etal_2014} defined the eigenvectors of $\bm{A}$ as follows:
\begin{equation}
    \bm{\mathit{\Phi}}=\bm{D}_{1}\widetilde{\bm{W}}\widetilde{\bm{\mathit{\Sigma}}}^{-1}\bm{V}\bm{\mathit{\Lambda}}^{-1},
    \label{exactvector}
\end{equation}
where the matrix $\bm{\mathit{\Phi}}$ consists of the eigenvectors of $\bm{A}$. The definition \eqref{exactvector} differs from that in the original DMD formulation \citep{Schmid_2010}, but Eq. \eqref{exactvector} guarantees that $\bm{\mathit{\Phi}}$ corresponds to the exact eigenvectors of $\bm{A}$ \citep{Brunton_Kutz_2019}. 

\section{Thermal cycle analysis considering the temperature dependent thermal property}
\label{kc_temp}
In Section ~\ref{thermal_cp_condition}, thermal conductivity $k$ and specific heat capacity $c$ are treated as constants independent of temperature. 
Here, we perform a thermal cycle analysis that accounts for these temperature dependencies and verify its validity by comparing the results.
According to \cite{Kim_1975}, the temperature dependence of $k$ and $c$ are expressed as listed on Table~\ref{kc_temperature}.

\begin{figure*}[t]   
  \centering
  \includegraphics[width=\textwidth]{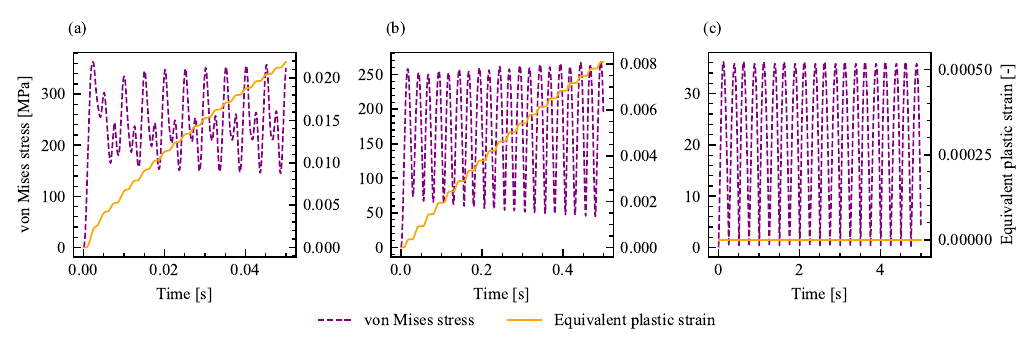}
  \caption{Time evolution of von Mises stress and equivalent plastic strain for (a)$200\,\mathrm{Hz}$, (b)$20\,\mathrm{Hz}$, and (c)$2\,\mathrm{Hz}$, considering temperature dependence of the thermal conductivity and specific heat.}
  \label{kc_temp_timehistory}
\end{figure*}

\begin{figure*}[t]   
  \centering
  \includegraphics[width=\textwidth]{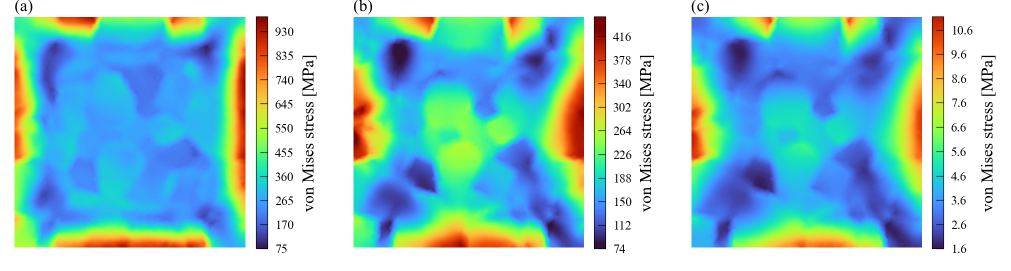}
  \caption{von Mises stress distributions obtained from thermal cycle analyses at the end of the 10th cycle for (a)$200\,\mathrm{Hz}$, (b)$20\,\mathrm{Hz}$, and (c)$2\,\mathrm{Hz}$, considering temperature dependence of the thermal conductivity and specific heat.}
  \label{kc_temp_vmheat}
\end{figure*}

The time histories of the spatially averaged von Mises stress and equivalent plastic strain, when the temperature dependence of thermal conductivity and specific heat is included, are shown in Fig.~\ref{kc_temp_timehistory} for (a) $200\,\mathrm{Hz}$, (b) $20\,\mathrm{Hz}$, and (c) $2\,\mathrm{Hz}$.  
Figure~\ref{kc_temp_vmheat} presents the spatial distributions of von Mises stress at the end of the 10th cycle under the same temperature-dependent conditions for (a) $200\,\mathrm{Hz}$, (b) $20\,\mathrm{Hz}$, and (c) $2\,\mathrm{Hz}$.

The results indicate that, although the absolute magnitudes of stress and strain are slightly affected by the temperature dependence, the spatial and temporal characteristics of the response remain consistent.
This behavior can be interpreted based on the Fourier number.
Although the change in the thermal cycle frequency changes the Fourier number by roughly one order of magnitude, the influence of temperature dependent thermal conductivity and specific heat is largely self-cancelling: the increase in thermal conductivity and the increase in specific heat from $300\,\mathrm{K}$ to $700\,\mathrm{K}$ approximately compensate each other, resulting in only about 1.3 times change in the thermal diffusivity and Fourier number.

\begin{table*}[htbp]
\centering\caption{Temperature dependent thermal properties}\label{kc_temperature}
\begin{tabular}{ccc}\hline
    Thermal property \citep{Kim_1975} & Value & Unit \\ 
    \hline
    Thermal conductivity, $k$ & $9.248+1.571\times 10^{-4}\theta(\mathrm{K})$ & $\si{\watt\metre^{-1}\kelvin^{-1}}$ \\
    Specific heat capacity, $c$ & $458.9848+13.28\times 10^{-2}\theta(\mathrm{K})$ & $\si{\joule\,\kilo\gram^{-1}\kelvin^{-1}}$\\
    \hline
\end{tabular}
\end{table*}

\section{Sensitivity analysis of the Taylor-Quinney parameter}
\label{sensitivity_chi}
In Section~\ref{thermal_cp_condition}, the Taylor-Quinney parameter $\chi$ is set to 0.85 and this value is used at all over the analysis domain, and throughout the simulation. 
This assumption is widely accepted in the thermal-crystal plasticity simulations previously conducted. 
However, the Taylor-Quinney parameter is reported to have dependencies on strain rate, and microstructure \citep{Soares_Hokka_2021}.
Furthermore, the self-heating caused by plastic work accumulates as the number of cycles increases, and may contribute to a rise in temperature.
To investigate the influence of the choice of Taylor-Quinney parameter on the simulation conducted in present study and the influence of the cumulative effect of self-heating over the range of thermal cycling conditions covered by this study, the thermal cycle simulations adapting $\chi=0$ are also conducted.
All settings are the same as those described in Section ~\ref{thermal_cp_condition}, except for the values of the Taylor-Quinney parameter.

\begin{figure*}[t]   
  \centering
  \includegraphics[width=\textwidth]{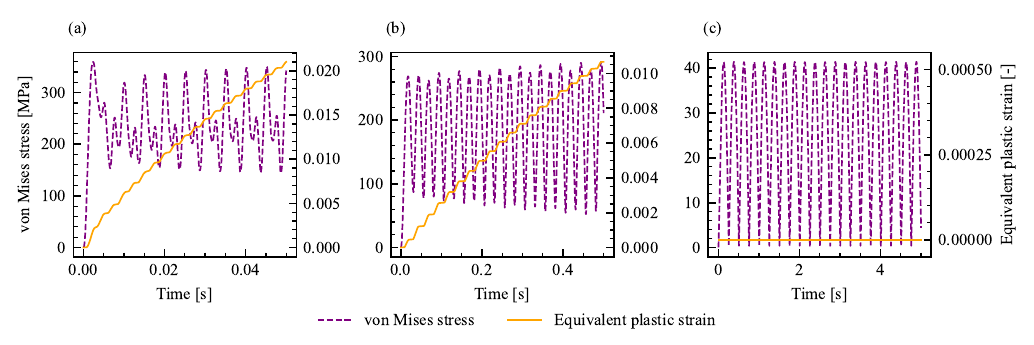}
  \caption{Time evolution of von Mises stress and equivalent plastic strain for (a)$200\,\mathrm{Hz}$, (b)$20\,\mathrm{Hz}$, and (c)$2\,\mathrm{Hz}$, adapting $\chi=0$.}
  \label{chi0_timehistory}
\end{figure*}

\begin{figure*}[t]   
  \centering
  \includegraphics[width=\textwidth]{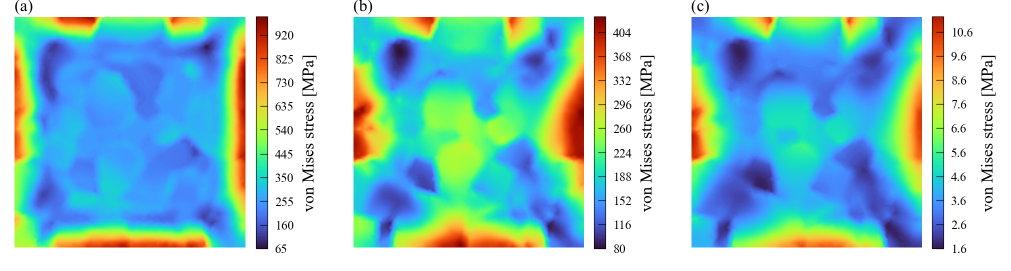}
  \caption{von Mises stress distributions obtained from thermal cycle analyses at the end of the 10th cycle for (a)$200\,\mathrm{Hz}$, (b)$20\,\mathrm{Hz}$, and (c)$2\,\mathrm{Hz}$, adapting $\chi=0$.}
  \label{chi0_vmheat}
\end{figure*}

As shown in Figures~\ref{chi0_timehistory} and \ref{chi0_vmheat}, the time evolution and spatial distribution of von Mises stress and equivalent plastic strain are nearly identical when $\chi=0$ compared to the case with $\chi=0.85$. 
This indicates that the heat generated by plastic deformation in the present simulations has a negligible effect on the temperature field relative to the imposed thermal cycle at the boundaries. 
Consequently, the choice of Taylor-Quinney parameter and the self-heating effect accumulated during the present thermal cycling conditions do not significantly affect the observed stress oscillations and inhomogeneity patterns in the present study.

\section{Temperature effect embedded in crystal plasticity model}
\label{thermal_softening}
The crystal plasticity model described in Section~\ref{thermocpmodel} incorporates temperature effects through the slip hardening law (Eq.~\eqref{sliphardeninglaw}) and temperature-dependent elastic coefficients. 
To validate the temperature dependence embedded in the model, uniaxial tensile simulations (Section~\ref{thermal_cp_condition}) were performed at an elevated temperature of $613\,\si{\kelvin}$.

\begin{figure}[t]
\begin{center} 
\includegraphics[width=0.45\textwidth]{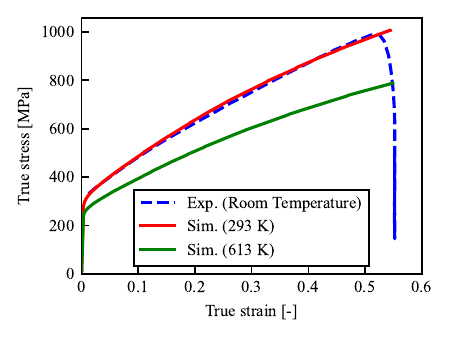}
\caption{Comparison of simulated and experimental stress-strain curves under tensile loading.} 
\label{thermal_softening_result} 
\end{center}
\end{figure}

Figure~\ref{thermal_softening_result} compares the experimental stress-strain curve at room temperature \citep{Yan_etal_2012} with the corresponding simulation result ($293\,\si{\kelvin}$), as well as the simulation result at $613\,\si{\kelvin}$. 
The decrease in 0.2\% proof stress due to thermal softening is approximately $45\,\si{\mega\pascal}$, which is reasonably consistent with the reported value of $74\,\si{\mega\pascal}$ \citep{Scherer_etal_2024}. 
Note that the reported value is obtained from a different sample than the one used for calibration of the present model. 

It should be noted that the temperature dependence of the yield stress exhibits considerable variation across the literature \citep{Pawel_etal_1996,Desu_etal_2016,Scherer_etal_2024}, reflecting its strong dependence on microstructural features such as dislocation density, grain size, and strain rate. 
These results confirm that the present model reasonably captures the effect of temperature on yield behavior within the expected range of variability.

\section{Thermal cycle and DMD analysis considering distributed voids and precipitates}
\label{void_particle}

In the main analysis of this study, thermal cycle responses were evaluated using homogeneous polycrystalline materials.
However, in actual materials, the spatial distribution of the stress field is believed to be influenced by the presence of microscopic defects such as voids and precipitates.
Therefore, in this appendix, as an application of this method, we present the thermal-mechanical responses and DMD analysis results obtained when these defects are introduced, and examine the applicability of the analysis framework used in this study.

\begin{figure}[t]
\begin{center} 
\includegraphics[width=0.45\textwidth]{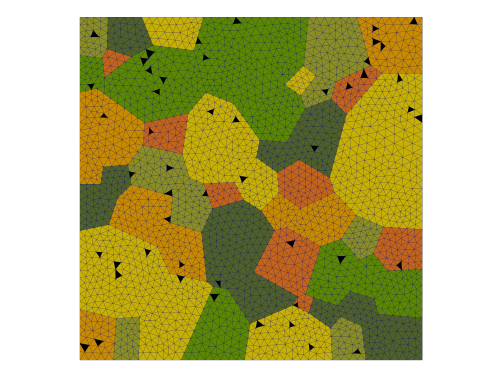}
\caption{Analysis model considering the distributed voids and particle.} 
\label{void_particle_model} 
\end{center}
\end{figure}

The analysis model in this appendix is created by using the same crystal structure and boundary conditions as described in the main text (Section \ref{thermal_cp_condition}), and defects are introduced by randomly selecting elements corresponding to approximately 1\% of the total.
The randomly selected elements are shown in black in Figure \ref{void_particle_model}.
For vacancies, the selected elements were removed, and the internal boundaries were treated as adiabatic boundaries.
On the other hand, for precipitates, the selected elements were assigned thermal and mechanical properties (Table \ref{particle_table}) based on $\mathrm{Cr}_{23}\mathrm{C}_{6}$.
All other conditions are identical to those in the main text.

\begin{table*}[htbp]
\centering\caption{Properties of $\mathrm{Cr}_{23}\mathrm{C}_{6}$}\label{particle_table}
\begin{tabular}{ccc}\hline
    Mechanical property \citep{Wu_etal_2022} & Value & Unit \\
    \hline
    Young's modulus & $348.8$ & $\si{\giga\pascal}$ \\
    Poisson's ratio & $0.300$ & $-$ \\
    \hline
    Thermal property \citep{Wu_etal_2022,Gong_etal_2020} & Value & Unit \\ 
    \hline
    Mass density, $\rho$ & $7168$ & $\si{\kilo\gram \metre^{-3}}$ \\
    Thermal conductivity, $k$ & $9.71$ & $\si{\watt\metre^{-1}\kelvin^{-1}}$ \\
    Specific heat capacity, $c$ & $21.735$ & $\si{\joule\,\kilo\gram^{-1}\kelvin^{-1}}$\\
    Thermal expansion coefficient, $\beta$ & $2.58\times 10^{-5}$ & $\si{\kelvin^{-1}}$\\
    \hline
\end{tabular}
\end{table*}

\begin{figure}[t]
\begin{center} 
\includegraphics[width=\textwidth]{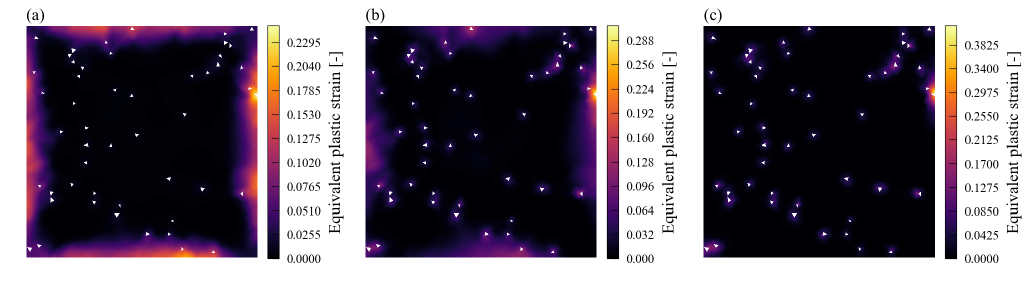}
\caption{Equivalent plastic strain distributions obtained from thermal cycle analyses at the end of the 10th cycle for (a)$200\,\mathrm{Hz}$, (b)$20\,\mathrm{Hz}$, and (c)$2\,\mathrm{Hz}$, under the presence of voids.} 
\label{void_eqps_heatmap} 
\end{center}
\end{figure}

First, we present the results obtained when voids were introduced.
Figure \ref{void_eqps_heatmap} shows the distribution of equivalent plastic strain at the end of 10 cycles.
Compared to the homogeneous material shown in Fig.~\ref{eqps_heatmap_10cycle}, significant localization of plastic strain is observed in the vicinity of the voids.
Even in the 2 Hz case, which exhibited a thermoelastic response for the homogeneous material, the presence of voids resulted in the accumulation of equivalent plastic strain.
Furthermore, at 200 Hz and 20 Hz, where the temperature gradient within the region is large, the presence of voids near the boundary, where the amplitude of temperature change is particularly large, strongly induces localized plastic deformation.

\begin{figure}[t]
\begin{center} 
\includegraphics[width=\textwidth]{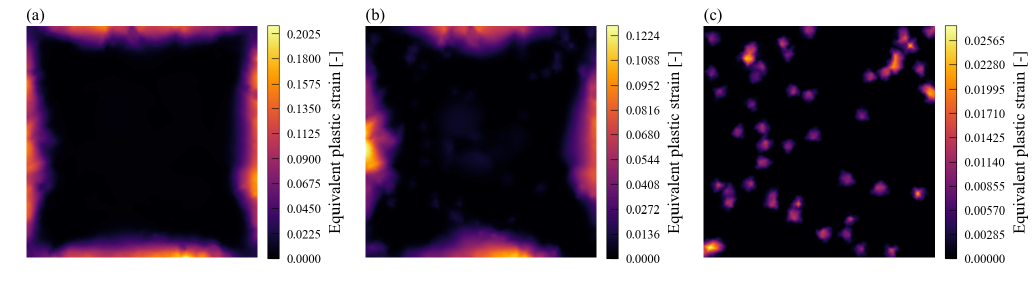}
\caption{Equivalent plastic strain distributions obtained from thermal cycle analyses at the end of the 10th cycle for (a)$200\,\mathrm{Hz}$, (b)$20\,\mathrm{Hz}$, and (c)$2\,\mathrm{Hz}$, under the presence of particle.} 
\label{particle_eqps_heatmap} 
\end{center}
\end{figure}

Next, we discuss the results obtained when inclusions are introduced.
As shown in Fig.~\ref{particle_eqps_heatmap}, a localized increase in plastic strain is observed around the inclusions; however, compared to the case with voids, the extent of this increase is limited.
On the other hand, due to the increased heterogeneity within the material, the spatial distribution of the stress and temperature fields tends to become more complex than in a homogeneous material.

As described above, the introduction of defects causes a distinct change in the thermal-mechanical response behavior, particularly by increasing the nonlinearity of the response through the manifestation of local plasticity.
Consequently, simple scaling laws that were effective for homogeneous materials may not be sufficient to adequately describe these responses.
Classical scaling laws based on the Fourier number predict the transition between quasi‑steady and non‑steady regimes only at a macroscopic level, whereas the present framework reveals how such regimes manifest as heterogeneous stress structures at the grain scale.

On the other hand, focusing on the results of the DMD analysis, it was confirmed that even when the response becomes complex in this manner as described below, the spatiotemporal evolution of the simulated stress field can still be represented by a limited number of DMD modes.
In other words, even when local nonlinearities or inhomogeneities increase, DMD functions as an effective method for extracting the dominant spatiotemporal structure of the analysis data.

\begin{figure*}[t]   
  \centering
  \includegraphics[width=\textwidth]{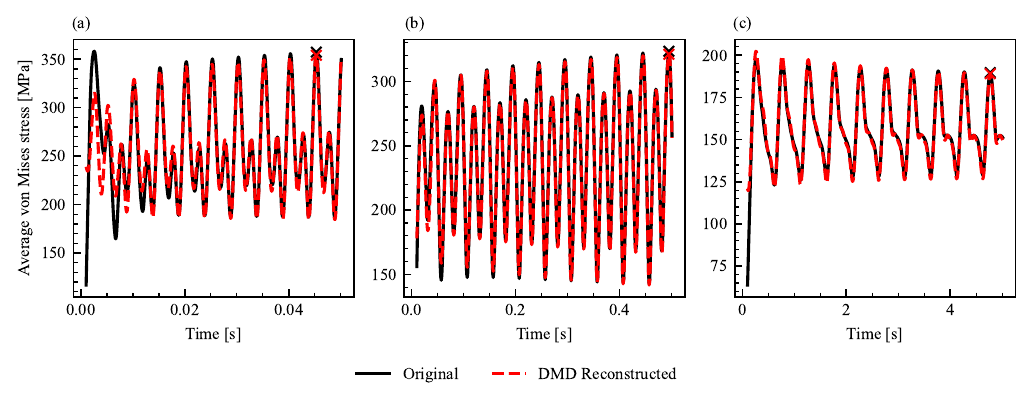}
  \caption{Comparison of the temporal evolution of the spatially averaged von Mises stress between the original snapshots and the DMD-reconstructed data for three datasets corresponding to thermal cycle frequencies of $200\,\mathrm{Hz}$, $20\,\mathrm{Hz}$, and $2\,\mathrm{Hz}$, under the presence of voids. DMD reconstruction employs the top 20 modes with Hankel dimension $h=15$. Cross markers indicate the time steps corresponding to the maximum stress in the final cycle, which are used in Fig.~\ref{void_dmd_vector_visualization} to visualize the distribution of DMD modes. Panels (a), (b), and (c) correspond to datasets with increasing thermal cycle frequency.}
  \label{void_recon_mean_vm}
\end{figure*}

\begin{figure*}[htbp]   
  \centering
  \includegraphics[width=\textwidth]{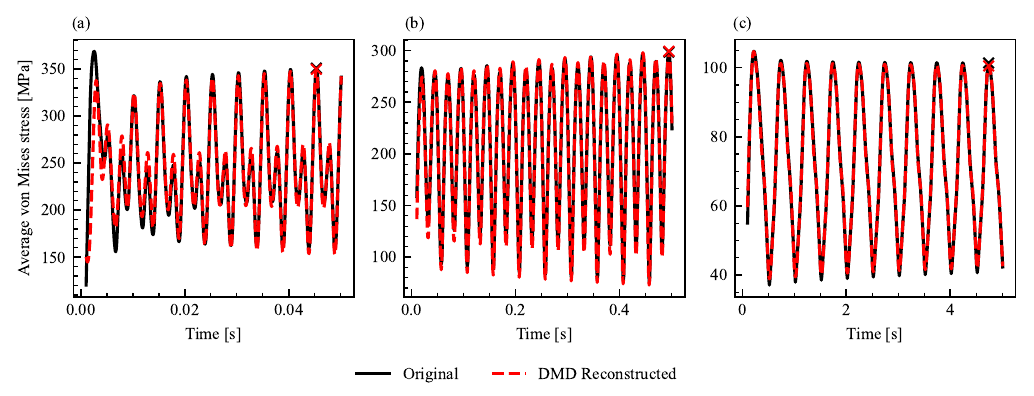}
  \caption{Comparison of the temporal evolution of the spatially averaged von Mises stress between the original snapshots and the DMD-reconstructed data for three datasets corresponding to thermal cycle frequencies of $200\,\mathrm{Hz}$, $20\,\mathrm{Hz}$, and $2\,\mathrm{Hz}$, under the presence of precipitates. DMD reconstruction employs the top 20 modes with Hankel dimension $h=15$. Cross markers indicate the time steps corresponding to the maximum stress in the final cycle, which are used in Fig.~\ref{particle_dmd_vector_visualization} to visualize the distribution of DMD modes. Panels (a), (b), and (c) correspond to datasets with increasing thermal cycle frequency.}
  \label{particle_recon_mean_vm}
\end{figure*}

Figures \ref{void_recon_mean_vm} and \ref{particle_recon_mean_vm} show the time-series plots of the mean von Mises stress (black lines) when voids and precipitates are taken into account, and the time-series plots of the mean values (red dotted lines) obtained from the von Mises stress field reconstructed by 20 Hankel DMD modes ($h=15$). 
The reconstruction errors, calculated using Eq.~\eqref{recon_error_def}, were approximately 13\% and 10\%, respectively. 
The increase in the reconstruction error when voids were considered is thought to be due to the fact that void regions in the data showing no stress oscillations at all appeared, making it difficult to represent the stress field using a superposition of vibration modes. This suggests that to improve DMD reconstruction errors for materials with voids, it would be effective to exclude the void regions from the DMD analysis domain in the first place.

\begin{sidewaysfigure*}   
  \centering
  \includegraphics[width=\textwidth]{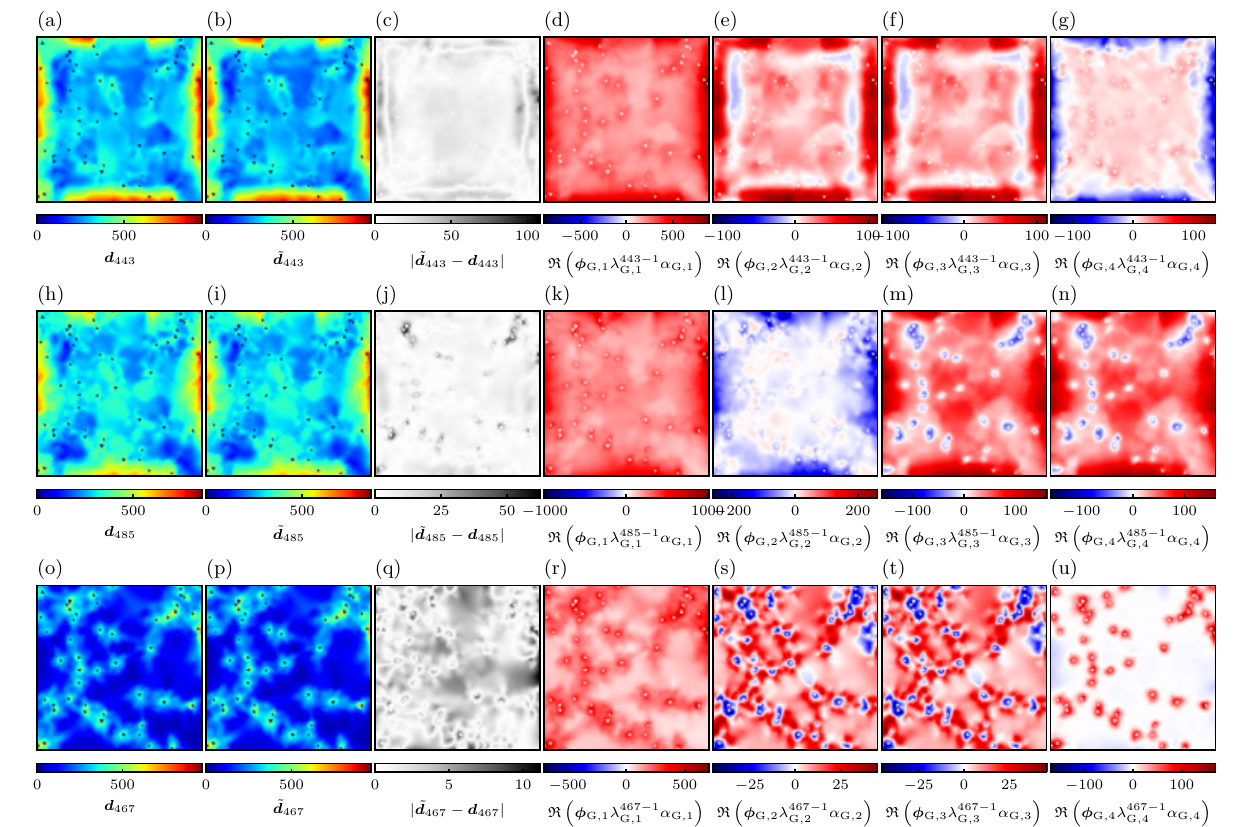}
  \caption{Spatial distribution of von Mises stress reconstructed by DMD for three datasets corresponding to thermal cycle frequencies of $200\,\mathrm{Hz}$, $20\,\mathrm{Hz}$, and $2\,\mathrm{Hz}$ (top to bottom), under the presence of voids, from left to right, each column shows: the original snapshot (Figure (a), (h), and (o)), the Hankel DMD ($h=15$) reconstruction using the top 20 modes (Figure (b), (i), and (p)) , the absolute reconstruction error (Figure (c), (j), and (q)), and the contributions of first four modes of greedy sorted DMD modes (Figure (d), (e), (f), (g), (k), (l), (m), (n), (r), (s), (t), and (u)). The snapshots correspond to the time of maximum stress in the final cycle, which is indicated by cross marker in Fig.~\ref{void_recon_mean_vm}.}
  \label{void_dmd_vector_visualization}
\end{sidewaysfigure*}

\begin{sidewaysfigure*}   
  \centering
  \includegraphics[width=\textwidth]{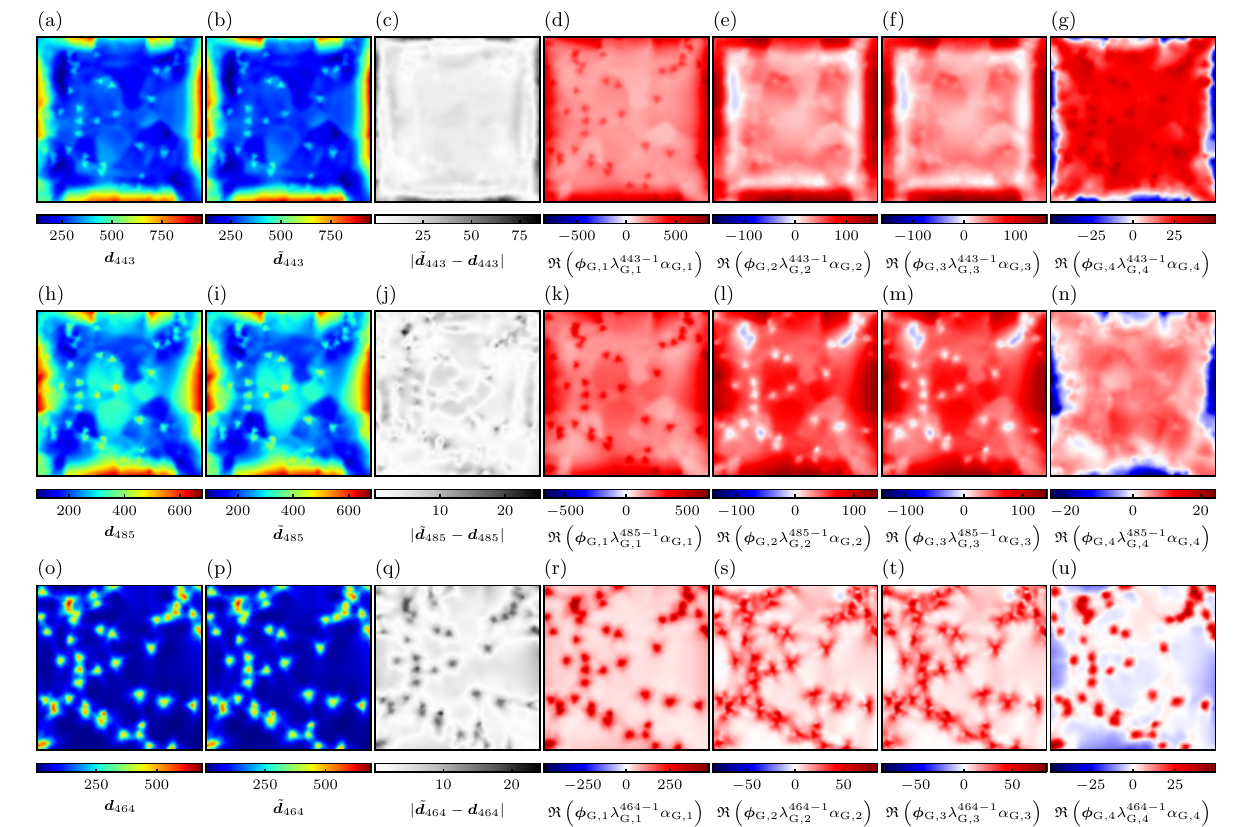}
  \caption{Spatial distribution of von Mises stress reconstructed by DMD for three datasets corresponding to thermal cycle frequencies of $200\,\mathrm{Hz}$, $20\,\mathrm{Hz}$, and $2\,\mathrm{Hz}$ (top to bottom), under the presence of precipitates, from left to right, each column shows: the original snapshot (Figure (a), (h), and (o)), the Hankel DMD ($h=15$) reconstruction using the top 20 modes (Figure (b), (i), and (p)) , the absolute reconstruction error (Figure (c), (j), and (q)), and the contributions of first four modes of greedy sorted DMD modes (Figure (d), (e), (f), (g), (k), (l), (m), (n), (r), (s), (t), and (u)). The snapshots correspond to the time of maximum stress in the final cycle, which is indicated by cross marker in Fig.~\ref{particle_recon_mean_vm}.}
  \label{particle_dmd_vector_visualization}
\end{sidewaysfigure*}

The stress field, reconstructed stress field, reconstruction error, and spatial distribution of the top four DMD modes at the time of the cross mark shown in Fig. \ref{void_recon_mean_vm} and Fig. \ref{particle_recon_mean_vm} are shown in Fig. \ref{void_dmd_vector_visualization} and Fig. \ref{particle_dmd_vector_visualization}, respectively.

Figures \ref{void_dmd_vector_visualization}(a), (h), and (o) and Figures \ref{particle_dmd_vector_visualization}(a), (h), and (o) show that, when voids or precipitates are present, the stress values increase around the voids and precipitates compared to the homogeneous material shown in Fig.~\ref{dmd_vector_visualization}.
Furthermore, the absolute values of the first-mode distribution (the 0 Hz mode for both voids and precipitates) are also higher than those in the homogeneous material. 
The degree of increase in both stress values and first-mode distribution is greater for voids than for precipitates.

The second and third modes at 2 Hz were observed in both the voids and precipitates. 
However, there was a difference in their spatial distribution: in Fig. \ref{void_dmd_vector_visualization} (s), (t), the phase of the vibration mode was inverted around the voids, whereas in Fig. \ref{particle_dmd_vector_visualization} (s), (t), no phase inversion occurred.

The difference in modal structure observed between voids and precipitates is thought to stem from differences in the manner in which heterogeneity is introduced. 
Voids introduce strong discontinuities in both stress and heat transport, producing an effect on the stress field that is similar to a change in boundary conditions. 
As a result, a time response different from that of the surrounding region is locally formed in the vicinity of the void, manifesting as a phase inversion of the vibration mode. 
On the other hand, while precipitates introduce heterogeneity in material properties, the response as a continuum is maintained; consequently, the stress field is merely spatially distorted, and the phase structure does not change significantly.

These results demonstrate that the CPFEM-DMD-based analysis framework proposed in this study is applicable not only to homogeneous materials but also to more complex microstructures containing voids and precipitates.
The differences in response caused by the introduction of voids and precipitates are clearly manifested as changes in the spatial distribution of modes and phase structure; the differences in spatial modes between voids and precipitates are not merely differences in distribution but can be interpreted as differences in the spatial synchronization of time responses. It has been demonstrated that DMD functions as a method for systematically capturing response differences caused by heterogeneity.
In other words, this method is suggested to be an effective tool for organizing and interpreting the structure of high-dimensional simulation data under a wide range of conditions, from thermoelastic to thermoelastoplastic responses.

\section{DMD applicability to non-periodic thermal input}
\label{appf}
To further assess the applicability of DMD beyond strictly periodic inputs, an additional analysis was conducted using a non-periodic thermal boundary condition. 
In this case, the temperature was increased monotonically from 300 K to 700 K over a period of $50\,\si{\milli s}$, and DMD was applied to the resulting stress field.
All other conditions are identical to those in the main text.

\begin{figure*}[t]   
  \centering
  \includegraphics[width=\textwidth]{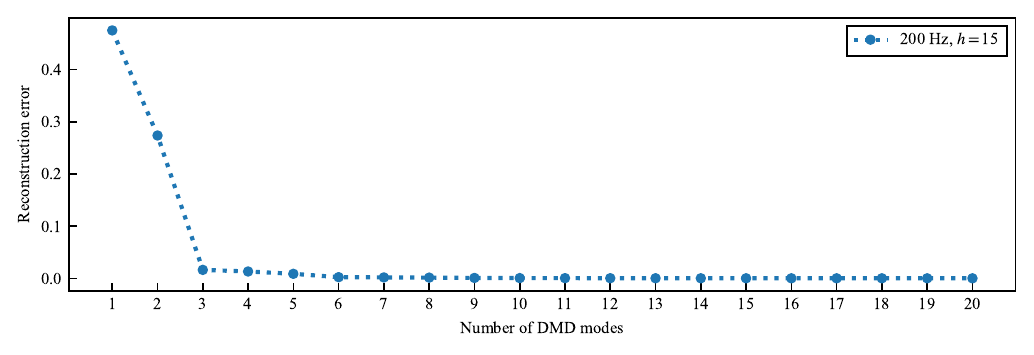}
  \caption{Reconstruction error versus the number of DMD modes for the monotonic thermal input. The results are shown for greedy-ordered DMD modes.}
  \label{reconstruction_error_monotonic}
\end{figure*}

Figure~\ref{reconstruction_error_monotonic} shows the reconstruction error as a function of the number of DMD modes. 
The reconstruction error decreases rapidly with increasing number of modes, reaching approximately $0.03\%$, indicating that the stress field can be accurately represented with a limited number of modes.

\begin{figure*}[t]   
  \centering
  \includegraphics[width=\textwidth]{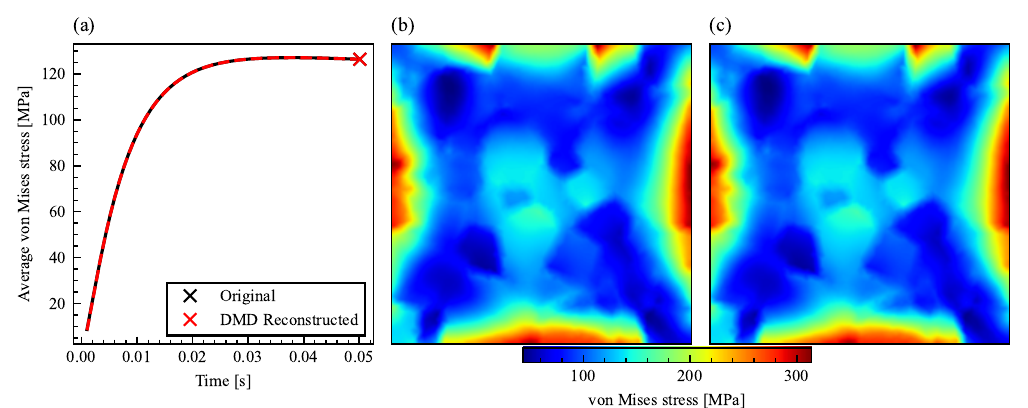}
  \caption{(a) Temporal evolution of the spatially averaged von Mises stress obtained from the thermal–crystal plasticity simulation and the DMD reconstruction under monotonic thermal input. 
(b) and (c) show the spatial distribution of the von Mises stress from the simulation and the DMD reconstruction, respectively, at the time indicated in (a).}
  \label{reconstruction_comparison_monotonic}
\end{figure*}

Figure~\ref{reconstruction_comparison_monotonic} compares the temporal evolution of the spatially averaged von Mises stress between the thermal–crystal plasticity simulation and the DMD-reconstructed data, along with representative spatial distributions at selected time steps. 
The DMD reconstruction shows excellent agreement with the original simulation results, both in terms of temporal evolution and spatial distribution.

These results indicate that DMD can effectively extract the dominant spatiotemporal structures even under non-periodic transient thermal loading. 
However, it should be noted that this validation is limited to the monotonic input considered here, and the applicability to more complex non-periodic conditions remains to be investigated.

\bibliographystyle{elsarticle-harv}
\bibliography{reference.bib}

\end{document}